\RequirePackage{ifpdf}
\documentclass[letterpaper]{JHEP3}
\usepackage{amsmath}
\usepackage{epsfig}
\usepackage{subfigure}
\usepackage{amsfonts}
\usepackage{amssymb}
\usepackage{epstopdf}

\newcommand{\mbb}{\mathbb}
\newcommand{\mc}{\mathcal}

\newcommand{\roughly}[1]{\mathrel{\raise.3ex\hbox{$#1$\kern-0.85em
\lower1ex\hbox{$\sim$}}}}

\newcommand{\lsim}{\roughly<}
\newcommand{\gsim}{\roughly>}

\def\mc{\mathcal}
\def\nn{\nonumber}

\newcommand{\be}{\begin{equation}}
\newcommand{\bee}{\begin{equation}}
\newcommand{\ee}{\end{equation}}
\newcommand{\beea}{\begin{eqnarray}}
\newcommand{\eea}{\end{eqnarray}}
\newcommand{\bea}{\begin{eqnarray}}

\def\vo{\mathcal{V}}

\def\pref#1{(\ref{#1})}

\def\cG{{\cal G}}

\def\cL{{\cal L}}

\def\cO{{\cal O}}

\def\cV{{\cal V}}

\newbox\charbox
\newbox\slabox
\def\slsh#1{{      
        \setbox\charbox=\hbox{$#1$}
        \setbox\slabox=\hbox{$/$}
        \dimen\charbox=\ht\slabox
        \advance\dimen\charbox by -\dp\slabox
        \advance\dimen\charbox by -\ht\charbox
        \advance\dimen\charbox by \dp\charbox
        \divide\dimen\charbox by 2
        \raise-\dimen\charbox\hbox to \wd\charbox{\hss/\hss}
        \llap{$#1$}
}}

\def\exd{{\hbox{d}}}

\def\cL{{\cal L}}

\def\cO{{\cal O}}
\def\cG{{\cal G}}

\def\bea{\begin{eqnarray}}
\def\eea{\end{eqnarray}}
\def\be{\begin{equation}}
\def\ee{\end{equation}}

\def\ssW{{\scriptscriptstyle W}}
\def\ssX{{\scriptscriptstyle X}}

\def\KK{{\scriptscriptstyle KK}}

\def\BD{{\scriptscriptstyle BD}}

\def\KK{{\scriptscriptstyle KK}}

\title{Robust Inflation from Fibrous Strings}

\author{C.P.~Burgess,${}^{1,2}$ M.~Cicoli,${}^{3,4,5}$ S. de Alwis ${}^6$ and F.~Quevedo${}^{5,7}$ \\ 

{\it 
${}^1$ Department of Physics \& Astronomy, McMaster University\\ \qquad 1280 Main Street West, Hamilton ON, Canada.\\
${}^2$ Perimeter Institute for Theoretical Physics\\
\qquad 31 Caroline Street North, Waterloo ON, Canada.\\
${}^3$ Dipartimento di Fisica e Astronomia, Universit\`a di Bologna,\\ $\qquad$ via Irnerio 46, 40126 Bologna, Italy.\\
${}^4$ INFN, Sezione di Bologna, Italy.\\
${}^5$ Abdus Salam ICTP, Strada Costiera 11, Trieste 34014, Italy.\\
${}^6$ Physics Department, University of Colorado, Boulder, CO 80309, USA.\\
${}^7$ DAMTP, University of Cambridge, Wilberforce Road,  Cambridge, CB3 0WA, UK.
}}

\preprint{DAMTP-2015-89}

\date{\today}

\abstract{Successful inflationary models should ($i$) describe the data well; ($ii$) arise generically from sensible UV completions; ($iii$) be insensitive to detailed fine-tunings of parameters and ($iv$) make interesting new predictions. We argue that a class of models with these properties is characterized by relatively simple potentials with a constant term and negative exponentials. We here continue earlier work exploring UV completions for these models --- including the key (though often ignored) issue of modulus stabilisation --- to assess the robustness of their predictions. We show that string models where the inflaton is a fibration modulus seem to be robust due to an effective rescaling symmetry, and fairly generic since most known Calabi-Yau manifolds are fibrations. This class of models is characterized by a generic relation between the tensor-to-scalar ratio $r$ and the spectral index $n_s$ of the form $r \propto (n_s-1)^2$ where the proportionality constant depends on the nature of the effects used to develop the inflationary potential and the topology of the internal space. In particular we find that the largest values of the tensor-to-scalar ratio that can be obtained by generalizing the original set-up are of order $r \lsim 0.01$. We contrast this general picture with specific popular models, such as the Starobinsky scenario and $\alpha$-attractors. Finally, we argue the self consistency of large-field inflationary models can strongly constrain non-supersymmetric inflationary mechanisms.}

\begin{document}

\section{Introduction}

The cumulative WMAP \cite{wmap} and Planck \cite{planck15} results raise the bar for those who profess to be able to read the cosmic tea leaves. It no longer suffices to write down simple potentials and get things roughly right since the data now discriminates more finely than just choosing concave-down from concave-up potentials. But neither is the data good enough to sift finely amongst all the extant theory proposals, so theorists must recalibrate the most fruitful ways to organize possible theoretical alternatives. 

In this paper we propose a set of desirable criteria on the theory side, and state the direction towards which we believe they suggest the present data are pointing. We believe these criteria are conservative, and indeed are already widely used amongst theorists when deciding how to organize what to think about. What is interesting is that these criteria already appear to help differentiate among many popular models, and we hope their enunciation can help observers who find themselves rummaging through bargain bins at their neighborhood theory store. 

A start is to organize inflaton potentials into categories based on theoretical inputs, so these inputs can be discriminated based on what observations seem to be telling us. Examples along these lines are quadratic potentials, $V = \frac12 \, m^2 \phi^2$, that capture fields deep within a potential well \cite{chaotic}, trigonometric potentials, $V = V_0\,[1 - \cos(\phi/f)]$, whose inflation \cite{NInf} captures axion-like models and exponential potentials, $V = V_0 - V_1 \, e^{- \phi/f}$, that often emerge as the low-energy limit of extra-dimensional moduli \cite{ExpPot, SInfPP, Cicoli:2011zz} (and more recently have been identified as particularly attractive descriptions of the data \cite{Attractors, NCNInf}).  

We evaluate inflationary classes of models like this according to how well they satisfy the following criteria:
\begin{enumerate}
\item \emph{Data fitting}: First and foremost they should agree with the data, but they should also do so robustly. That is, the agreement between theory and experiment should overlap strongly rather than tangentially. One way to quantify this criterion formally is through Bayesian comparisons with the data, such as those of \cite{EInf}. Although one can argue about the priors used, such a Bayesian comparison differentially punishes models that stray from experimentally favoured regions as parameters are varied.

\item \emph{UV completion}: Second, they should plausibly embed into some sort of UV completion.\footnote{We do not mean here to slavishly believe all of the details of any particular UV completion, which might be quite baroque. The point is rather that it is important that UV completions for the class of interest exist, and whether they restrict the parameter range of the class in an interesting way.} This criterion expresses a very basic fact: Nature is a whole that does not subdivide itself according to the academic disciplines. Any model that successfully describes cosmology must also play nicely with whatever else we know about Nature at the relevant energies. This is harder than it looks for inflationary models because these by assumption are in a regime where quantum and gravitational effects are both in play in an observable way (usually because the inflationary scale is at such very high energies). The model should therefore embed into a UV framework wherein the myriad corrections (both quantum and gravitational) to the simplest picture are under quantitative control. 

\item \emph{Naturalness}: Third, the successful class of models should embed into the UV completion in a technically natural way. That is, any choices made for cosmology should have roughly the same form within the effective theory relevant at {\em any} scale we choose. In practice this asks that any states at very high energies not dramatically ruin the choices made for effective parameters once they are integrated out (see, {\em e.g.} \cite{CCrev} for a more precise statement of what technical naturalness asks). 

\item \emph{New predictions}: A bonus is if the class of interest also makes a new or specific prediction for the size of a hitherto unmeasured effect (such as for the tensor-to-scalar ratio, $r$, say) that can be further tested.
\end{enumerate}
Although items 1 and 4 are not controversial, several comments are in order about 2 and 3; both about why they should be true and why (even if true) they are useful. 

First, we remark that although in principle criterion 2 can be done with any theory of quantum gravity as UV completion, the present state of the art seems only to allow this to be done with sufficient precision using string theory. In practice embedding into string theory is what we ask. 

Second, one might worry that criterion 2 is pretty ineffective inasmuch as it does not much restrict one's choices. After all, can't one get {\em anything} from string theory (or one's favourite alternative)?  The challenge here really has several parts, depending on the cosmological model of interest. In contrast to models ({\em e.g.} bouncing cosmologies) that rely on gravity in a strongly quantum regime, string theory provides a precise framework within which to sort out whether controlled predictions can be made at all. 

Furthermore, embedding into string theory doesn't just mean using fields that might plausibly arise in some sort of stringy context. It asks the embedding to be done at a level of control that includes {\em all} contributions that are as large as the one desired. For inflationary models the hard part about cosmology usually is to compute reliably the scalar potential of interest; in particular to stabilize the system's various moduli. It is, after all, a waste of time to find a shallow direction in some potential if there are other, ignored, field directions where the potential is much steeper. The good news is that this is even possible, since tools for doing so \cite{GKP, KKLT, KKLMMT, LVS} have been known for more than a decade. But experience with these tools also tell us that if you are not stabilizing all fields --- even those not directly involved in the cosmology of interest -- then you have not yet gotten to the hard part of the problem. 

Even so, one might also think that a stringy provenance can be found for any possible cosmological option. Although this needs reassessing as more is learned, the same has so far not proven to be true in particle physics applications of string theory.\footnote{For instance, although field representations can be chosen quite freely in particle models, strings seem only to give small-dimensional representations.} It is still early days for string cosmology, of course, and it is true that not all inflationary options have yet emerged from string theory \cite{SInfPP}. Yet all the options found so far seem to have difficulty getting a large tensor-to-scalar ratio, $r$, for example -- with the models of \cite{eva} so far providing a high-water mark.

Finally, criterion 3 is the flip-side of `decoupling,' {\em i.e.}~the usual belief that nothing much at low energies should depend on what goes on at very high energies (which in practice is why science makes any headway at all). Effective field theories (EFTs) express this basic fact (and this is why they are useful), but also teach us that there are usually a few kinds of interactions that are more sensitive than most to the details of high-energy physics. Happily a successful description of cosmology seems to hinge on several of these: relatively small scalar masses and vacuum energies.

Criterion 3 would be a fairly obvious one if it were not that similar arguments applied to the cosmological constant and to particle physics at the LHC have not so far borne fruit. It is an open question whether this requires a rethink of naturalness but it is also true that news of its death is premature, at least until an equally clear alternative quantitative framework is available.\footnote{Possibly anthropic arguments ultimately will fill this role, but we believe it is too soon to tell.}

As argued elsewhere \cite{SInfPP, Cicoli:2011zz, NCNInf} we think there is a well-motivated class of models that does satisfy all of these criteria: the class of exponential potentials: $V = V_0 - V_1 \, e^{-\phi/f}$. They fare well in data comparisons \cite{EInf}; they are known often to emerge generically from extra-dimensional models with geometric moduli as inflaton \cite{ExpPot} and this survives more explicit embedding into string theory with modulus stabilization \cite{Fibre, kahlerinflation, polyinstanton}. The inflaton mass can be protected by symmetries (non-compact Abelian rescaling symmetries) in the same way as for axions (compact Abelian shift symmetries) \cite{NCNInf}, with the bonus that the corresponding `decay constant' is naturally of order $M_p$ (rather than $M_s \ll M_p$, as for axions). Finally, the parameter $f$ relates $r$ to $n_s$ by:
\be 
 r = 2\left(\frac{f}{M_p}\right)^2 (n_s - 1)^2 \,,
\label{rvsns}
\ee
where explicit UV completions give $f$ in a relatively narrow range around $M_p$ and so at face value suggest $r$ should be expected to be of order $10^{-3}$. These models are equivalent, after field redefinition, to the `$\alpha$-attractor' models of \cite{Attractors} with $\alpha = \frac 23 \left(\frac{f}{M_p}\right)^2$.

Yet the criteria are not vacuous, since for instance quadratic and trigonometric potentials no longer do well with point 1. And many well-known models that satisfy 1 do not seem to satisfy criterion 2 and/or 3, including (but not limited to) the popular Higgs Inflation \cite{HI} and curvature-squared models\footnote{UV completions for these models may yet emerge, and we believe this is more likely to happen if they are regarded as special cases of exponential potentials.} \cite{Starobinsky:1980te}. As we argue below, we also believe criterion 3 is a challenge for some versions of axion monodromy \cite{eva} that stray too far from a supersymmetric limit at high energies.

The remainder of the paper elaborates these points in several ways. \S\ref{sec:robust} starts by examining criterion 3 in more detail. In particular it explores how criterion 3 includes several independent issues, only one of which is the small inflaton mass (or $\eta$-problem). This section argues why criterion 3 applied to the vacuum energy is a strong condition that (in our present understanding) broadly disfavours models without approximate supersymmetry during inflation.  As applied to the inflaton mass \S\ref{sec:robust} briefly reviews the potential importance of pseudo-Goldstone bosons (pGBs) \cite{pGB, pGBrev} to inflation \cite{NInf}, and compares the merits of the trigononmetric potentials of compact, axion-like pGBs of \cite{NInf} with the exponential potentials of non-compact pGBs \cite{NCNInf, GonLin, CKST}. 

Finally, \S\ref{sec:GFibInf} changes gears and revisits the most concrete UV completion, Fibre Inflation \cite{Fibre}, for exponential models, in order to establish how robust it is to perturbations and how broadly the parameter $f$ can be varied. Although we do find many ways to perturb the model, none pushes the upper limit for $r$ larger than about $0.01$. We find the Fibre Inflation scenario is more robust than originally thought in several ways:
\begin{enumerate}
\item First, the existence of fibre moduli turns out to be very generic in that the overwhelming majority of Calabi-Yau (CY) manifolds are fibrations  \cite{GenericFibre}. This makes fibre K\"ahler moduli a very generic class in which to seek an inflaton. 

\item Second, models more general than the simplest model of Fibre Inflation turn out to give rise to similar physical implications. Even models without fibre moduli but with at least one modulus other than the overall extra-dimensional volume, may give rise to an inflationary scenario with similar properties as the original Fibre Inflation model \cite{Fibre}.

\item Finally, although the explicit potential is notoriously difficult to compute in Fibre Inflation (it involves string loops computed for a non-trivial compactified geometry) the inflationary consequences depend only on two very robust features. The first is the overall scaling of the potential with the extra-dimensional volume, $\vo$, and the second is the exponential form taken by the potential at large fields. The first suffices to show why the fibre moduli are always lighter than generic moduli, and why their mass (at the potential's minimum) is generically of order the Hubble scale during inflation. The exponential form at large field then shows why the fibre-modulus mass becomes much smaller than $H$ in the large-field inflationary regime. 
\end{enumerate} 

After presenting our conclusions in \S\ref{Concl}, in Appendix \ref{AppA} we contrast these features with the popular Starobinsky scenario \cite{Starobinsky:1980te} which has very similar observational properties but no known UV provenance. We examine this scenario from a string perspective, and explore how robust its effective field theory is to plausible corrections at low energies. The purpose is not to single out this particular model but to illustrate the line of thought one might try to pursue for any of the classes of inflationary models presently on the table. In Appendix \ref{AppB} we discuss a possible Calabi-Yau deformation of the toroidal orientifold model where string loop corrections have been computed \cite{BHK}. This model has a similar behaviour as the generic fibre case but shows a larger value of the effective `decay constant' $f$ in (\ref{rvsns}) leading to a larger prediction for the tensor-to-scalar ratio $r$ of order $0.01$.

\section{Robustness to UV effects}
\label{sec:robust}

From the point of view of microscopic physics the most revealing robustness constraints ask how the existence of other high-energy states can alter the basic inflationary picture. This section summarizes the usual ways this can happen, how models avoid these problems, then closes with a discussion of the implications of these considerations when the UV completion is a (large-volume) string model. Those familiar with these issues can be forgiven for jumping ahead to \S\ref{ssec:GenConstraints}, where they are used to draw a few generic constraints on the extra-dimensional volume in the case the UV completion is an extra-dimensional model.

\subsection{Generic constraints}

The basic observation is that weakly coupled quantum fluctuations of fields with mass $M$ contribute to effective couplings in the low-energy theory like $M$ raised to the appropriate dimension. (String physics tends to cap these contributions at the string scale, $M_s$.) This is considered to be problematic (or `unnatural' -- see {\em e.g.} \cite{CCrev} for the argument why) when these contributions are many orders of magnitude larger than the coupling's desired (or observed) value, because this involves a detailed cancellation between the contributions of physics at differing UV scales in a way not seen for any other hierarchies we understand. Technical naturalness is the statement that no such cancellations are required as particles of successive mass are integrated out. The problem is clearly most severe when couplings with positive dimension receive contributions from the heaviest virtual states, since the dangerous corrections are then amplified by positive powers of any large scale $M$. 

For inflation the scales of interest are the inflationary energy density, $V \sim M_{\rm inf}^4$, (related to the inflationary Hubble scale by $H \sim M_{\rm inf}^2/M_p$) and the inflaton mass, $m_\phi^2 \sim V''$. Problems arise because inflation requires these to be much smaller than the various UV scales. For higher-dimensional physics the UV scale is at most the string scale, $M_s$, since field-theoretic calculations fail at or below this point. For the 4D effective field theory (EFT) UV physics must similarly intrude at or below the Kaluza-Klein (KK) scale $M_\KK$. 

\subsubsection*{Vacuum energies}

In the effective theory it is the vacuum-energy term, $\cL = - V_0 \, \sqrt{-g}$, that has the coupling with the most positive dimension. In four dimensions $V_0 \sim M_{\rm inf}^4$ during inflation and absent a naturalness mechanism (like supersymmetry), the existence of large UV scales imposes two almost contradictory naturalness conditions on $M_{\rm inf}$. 

On one hand, the absence of supersymmetry (or any other way found to suppress how quantum fluctuations appear in the vacuum energy) implies quantum corrections to $V_0$ from UV physics at scale $M$ are generically of order $M^4/(4\pi)^2$, and so technical naturalness would imply $M_{\rm inf} \gsim M/\sqrt{4\pi}$, where $M \sim M_\KK$ (for a 4D EFT) or $M \sim M_s$ for a higher-dimensional EFT. 

On the other hand performing an inflationary analysis within the low-energy theory generically also requires the scale of inflation to be below the UV scale, $M_{\rm inf} \lsim M$. For instance, an inflationary scale as large as $M_{\rm inf} \sim M_s$ would preclude using ordinary field theory to infer whether inflation has occurred (as is usually done). The stronger condition $M_{\rm inf} \ll M_\KK$ is usually applicable in practice because most analyses are done within a 4D EFT instead of working with the full higher-dimensional field equations (see however \cite{HiDInf} for a fully extra-dimensional inflationary solution, including modulus stabilization). 

Strictly speaking, this last limit has a loophole: In very adiabatic settings a low-energy effective field theory need not have $V_0$ smaller than the UV scale, since physics with larger $V_0$ actually can be consistently described in the low-energy theory provided the energy within $V_0$ cannot be extracted.\footnote{Evolution is not sufficiently adiabatic -- more about which below -- for an effective theory if $H > M$ (and similarly for other time derivatives, like $\dot \phi/\phi$) \cite{NonAdiabEFT}, so (as is well known) one has no option to taking these smaller than UV scales.} However this is a fairly special situation and it is more generic to have $V_0$ smaller than the UV scale. The stronger condition would be required in particular if the EFT is to be valid both during inflation and during reheating, say, since the energy density in the potential is then mined to provide the initial heating of the Hot Big Bang. This is usually true even though Hubble friction acts to lower the inflaton energy between inflation and reheating.\footnote{One might hope Hubble friction after inflation might reduce the total energy available for reheating below the amount available in the inflationary potential, thereby allowing reheating to be understood in the EFT even if $M < M_{\rm inf}$. But in practice this does not really help much. For instance in a single-field model where the inflaton kinetic energy carries inflationary energy into the later Universe, energy conservation states $\partial_t\left( \frac12 \, \dot{\phi}^2 + V(\phi) \right) = -3H \dot{\phi}^{2}$, so the inflaton's work-energy theorem is:
\be
\frac12 \left( \dot{\phi_f}^{2} - \dot \phi_i^2 \right) = V(\phi_i) - V(\phi_f) -3\int_i^f \frac{\exd a}{a} \left( \frac{\dot{\phi}^{2}}{2} \right) 
= V(\phi_i) - V(\phi_f) -3 \ln \left( \frac{a_f}{a_i} \right)  \left\langle \frac{\dot{\phi}^{2}}{2} \right\rangle \,,\nn
\ee
where $\langle \cdots \rangle$ denotes the time average between $t_i$ and $t_f$. Unless $\ln (a_f/a_i)$ dominates the 60 inflationary $e$-folds usually required, the kinetic energy at reheating is not that suppressed relative to the inflationary potential.} 

Although it is a model-dependent issue how (and whether) reheating works in a given inflationary scenario (and so whether the EFT being used for inflation need be trusted to describe the energy extraction during reheating), the vacuum energy suggests the inflationary scale, $M_{\rm inf}$, cannot be too much below the UV scale within the domain of validity of the EFT. The same conclusion does not hold for supersymmetric vacua, provided the supersymmetry-breaking scale $\sqrt{F}=\sqrt{m_{3/2}M_p}$ (where $m_{3/2}$ is the gravitino mass) is much smaller than the UV scale $M$. The lower bound is then relaxed because supersymmetry can ensure the corrections to $V_0$ are instead naturally of order $m_{3/2}^2 M^2/(4\pi)^2$. 

In the extra-dimensional case it is considerations such as these (together with the other restrictions on corrections that supersymmetry often gives) that steer people towards supersymmetric constructions.

\subsubsection*{Scalar masses}

The coupling next-most sensitive to UV effects is usually a scalar mass, and this leads to the traditional UV $\eta$-problem faced by most inflationary models. This problem asks why the inflaton mass is so much smaller than the UV scale, given that slow roll requires it to satisfy $|\eta| \sim m_\phi^2/H^2 \ll 1$. Because $H \sim M_{\rm inf}^2/M_p \ll M_{\rm inf}$ this is a much stronger condition than the requirement $M_{\rm inf}$ be much smaller than the UV scale. 

The main difference with the vacuum energy is the existence of a new way to protect against UV corrections: by making the inflaton a (pseudo-) Goldstone boson \cite{pGB} for an (approximate) global symmetry\footnote{One might wonder how global symmetries can be consistent with UV completions given (say) the no-go theorems \cite{GSnogo, GSgo} for global symmetries in string theory. Although true, these theorems have loopholes like approximate accidental global symmetries (such as classical scale invariance) in the low-energy EFT or very weakly coupled gauge symmetries \cite{GSgo}.} \cite{pGBrev}. Goldstone bosons transform under the corresponding broken symmetry inhomogeneously $\delta \phi^a = \omega^a + \cdots$, where $\omega^a$ denote the broken group transformation parameter and ellipses denote possible terms involving the $\phi^a$'s. This shift is a defining feature because it indicates that the vacuum cannot be invariant under the corresponding symmetry, and if such a symmetry is exact precludes the appearance of the corresponding $\phi^a$ in the scalar potential, $V$. This is no longer strictly true if the symmetry is only approximate, but the dependence of $V$ on $\phi^a$ is then suppressed by the small symmetry breaking parameters (whose existence makes the symmetry approximate), and so is potentially much smaller than a generic size. The same suppression holds for the contributions to their mass coming from loops involving UV states {\em if} the approximate symmetry is extended to these states as well.\footnote{It is not unusual to hear that inflationary models are technically natural because the slow roll ensures the interactions within the inflaton potential are small, leading to an approximate shift symmetry in their absence. This shifty claim is not false but also not that useful since inflaton self-interactions are not the dangerous ones from the UV point of view. It is loops of very heavy states coupled to the inflaton that are dangerous, and the issue is whether {\em these} couplings enjoy any sort of approximate shift symmetry.}

This mechanism has made axions popular as inflatons, starting with \cite{NInf}, typically leading to trigonometric potentials for the breaking of the simplest compact groups. Less well studied are the exponential potentials that arise from the same arguments when applied to approximate noncompact symmetries \cite{NCNInf, CKST}, which are equally protected and turn out to provide better descriptions of post-Planck precision CMB measurements \cite{EInf}. It is these kinds of potentials that arise within the fibre class of string models described here, for which it is underlying symmetries of the moduli space (and sometimes extra-dimensional symmetries) that play this role, and and it is this underlying structure that helps protect the robustness of the inflationary predictions of these models \cite{SInfPP}. 

\subsubsection*{Nominally irrelevant interactions}

Most other effective interactions are irrelevant (in the technical sense) inasmuch as they are suppressed by UV scales rather than enhanced by them. This does not make them irrelevant (in the colloquial sense) to inflation, however, since the need for a teeny inflaton mass smaller than $H$ means that even normally neglected Planck-suppressed operators --- or Planck `slop' --- can contribute in a dangerous way. For instance once a near-constant potential, $V_0$, is allowed in the effective lagrangian (as inflationary models usually require) one must worry about the presence of the Planck-suppressed interaction: 
\be \label{PlanckSlop}
 \Delta \cL \propto {V_0 \; \phi^2}/{M_p^2} \,, 
\ee
since this contributes to the inflaton mass an amount $\delta m_\phi^2 \sim V_0/M_p^2 \sim H^2$ \cite{LiamEva}. 

The good news here is that most effects of much heavier UV physics can be integrated out before discussing inflation, leaving them to contribute largely through marginal or mass-suppressed interactions. Since most of these interactions are too small to be observable, most effects of heavy particles in the UV sector are not important \cite{NonAdiabExamples, InfEFT, HoverM}. The bad news is there are a few interactions like \pref{PlanckSlop} that cannot be neglected. Unlike for the previous naturalness problems --- that are so severe that they usually require a symmetry mechanism in the low-energy EFT --- it is generically not possible to decide whether dangerous Planck slop is absent {\em without} access to a UV completion of one form or another.  

A different way UV physics can intrude into inflationary predictions is if the physics involved is not adiabatic, since this also precludes its description in terms of a low-energy EFT \cite{NonAdiabEFT}. Although many examples illustrate this effect using Lorentz-breaking modifications to heavy-particle dispersion relations \cite{DispRel}, this Lorentz breaking is not a necessary feature since observable non-adiabatic effects (at least near horizon exit) are also known to be able to modify inflationary predictions \cite{NonAdiabExamples}. 

Ultimately, the absence of these effects is an issue of initial conditions. Here the good news is that inflationary expansion tends to iron away initial field motions, so the longer the inflationary epoch the fewer non-adiabatic motions remain available. But it is always possible (though not generic) to arrange an initial state that is a metastable `bomb' inasmuch as it initially hides higher-than inflaton energy density only to dump it at later times non-adiabatically onto an inflationary scenario. Part of the robustness of the models of interest here relies on the assumption (shared also by essentially all other models) that such a special initial state is not prepared.

\subsubsection*{Large field excursions}

A further complicating feature when discussing interactions within inflationary EFTs is often the need to discuss the motion of a canonically normalized field over Planckian distances in field space, $\Delta \phi \sim M_p$, such as is usually required \cite{Lyth} in models with observably large tensor-to-scalar ratio, $r$. Exploration of such a large field range can be safely done within a low-energy EFT provided only that large fields do not come with too large an energy cost, but it is of course never a good approximation to explore the EFT by expanding in powers of $\phi$. 

String theory shows that it is very easy to have large fields in a controlled low-energy approximation, and moduli are among the simplest examples of this. For supersymmetric configurations the energy does not depend at all on the value of the modulus field, so large fields are perfectly consistent with the low-energy limit. Typically they first appear in the kinetic term in the form
\be \label{SigMod}
 \cL_{\rm kin} = - \sqrt{-g} \; \frac12 \, \cG_{ab}(\phi) \, \partial_\mu \phi^a \, \partial^\mu \phi^b \,,
\ee
where $\cG_{ab}(\phi)$ is a metric for the target space within which $\phi^a(x)$ takes its values. The beauty of \pref{SigMod} is that $\cG_{ab}$ transforms like a tensor under field redefinitions $\phi^a \to \xi^a(\phi)$, so lends itself to expressing physical quantities in a field-redefinition independent way. Often symmetries dictate $\cG_{ab}$ up to normalization (or up to a few parameters), such as when $\phi^a$ are Goldstone bosons (for which $\cG_{ab}$ is a $G$-invariant metric on the target space, which is the coset $G/H$ when a symmetry group $G$ breaks to an unbroken subgroup $H$). In such cases $\cG_{ab}$ can be written explicitly without needing to rely on an expansion in powers of $\phi^a$. 

The covariance of this formulation also emphasizes how field redefinitions can be done to change large fields into small ones (and vice versa). In this language canonical normalisation amounts to choosing a Cartesian target metric, $\cG_{ab} \propto \delta_{ab}$ (which can always be done locally, though not globally unless the target space is flat). In general, if $\phi^a$ describe a range of field space where distances measured with $\cG_{ab}$ can diverge, we can always map to $\psi^a(\phi)$ where $\psi^a$ runs over a finite range. Once this is done the new metric, $\cG_{ab}(\psi)$, is generically singular near some points and this is how the theory remembers the infinite field range available to $\phi^a$.  

More generally, string theory is full of examples of large fields whose energies are not exactly zero, but are nevertheless small. A commonly encountered example is an extra-dimensional radius, $\rho$, for part of the extra-dimensional geometry. In this case field theoretic calculations require $\rho$ is much larger than the string length, $\ell_s$, and so the potential can be fruitfully expanded in powers of $\ell_s/\rho$: $V(\rho) \simeq V_0 + V_1 (\ell_s/\rho) + \cdots$. It is often the case that the distance to $\rho \to \infty$ is infinite, measured with the target-space metric, yet the large-field regime in this case can be the {\em only} place where the usual field-theoretic calculational tools work. (While having large fields is easy, what is usually hard\footnote{What makes this hard is the need for a convincing model to stabilize all moduli, since inflation requires knowing there are {\em no} directions in field space steeper than the desired inflationary trajectory. Progress on this point in inflationary string theory began with \cite{GKP, KKLMMT}. Extra-dimensional inflationary models built from moduli that do not address modulus stabilization have not yet gotten to the difficult part of the problem.} in string models is to get inflation; that is, it is hard to obtain a constant part of the potential like\footnote{This is what usually makes the overall volume and the dilaton not appropriate  inflaton candidates. Even though they both have a natural trans-Planckian domain they appear explicitly in all terms in the scalar potential through terms like the $e^K$ overall factor of the F-term potential and therefore it is difficult to  have a $V_0$ term. However all other combinations of K\"ahler moduli that do not appear in $K$ at leading order and are good starting points for inflaton candidates.} $V_0$.)

It is not unusual (though not inevitable) to find exponential potentials in this kind of construction \cite{ExpPot, NCNInf, SInfPP}. For instance compactifications for simple geometries like spheres give kinetic terms of radii of the form $\frac12 \, f^2 (\partial \rho)^2/\rho^2$ with $f = k M_p$ where $k$ is an order-unity constant, making the canonical variable 
\be
\phi = f  \ln \left( \frac{\rho}{\ell_s} \right) \,,
\ee
and $V(\rho(\phi))$ into a series in powers of exponentials, $e^{- \phi/f}$. Clearly for such a field $\phi \gg f \sim M_p$ is precisely the regime where the potential is under control (since then $\rho \gg \ell_s$). Notice also that because their kinetic terms arise as part of the higher-dimensional Eintein-Hilbert term $M_p$ is the natural choice for kinetic normalisation for these moduli, unlike for axions in string theory (for which the natural choice is $M_s$). This gives these moduli a leg up over axions when seeking trans-Planckian field displacements, since trans-Planckian motion for $\phi$ just corresponds to $\rho$ moving over many string lengths.

Of course the freedom to redefine fields always allows a large-$\phi$ choice for $V(\phi)$ to be translated into a choice for a small-$\rho$ singularity in the target-space metric, $\cG_{ab}$, allowing the one-parameter family of asymptotic exponential potentials to be recast as a one-parameter family of singularities near $\rho=0$, as is done {\em e.g.}~in the $\alpha$-attractor formalism of \cite{Attractors}.

\subsection{Applications to extra-dimensional models}
\label{ssec:GenConstraints}

We next turn to what some of the above naturalness constraints imply for generic extra-dimensional (including string) models. Although these arguments are not restricted to string theory, for later purposes it is useful to cast the role of UV scales in terms of string parameters like the string coupling, $g_s$, and squared string length $\alpha' \sim \ell_s^2 = M_s^{-2}$. For 4D applications in unwarped compactifications these are related to the Planck and KK scales by:
\be
 M_s\simeq \frac{g_s^{1/4} M_p}{\sqrt{\vo}}  \quad \hbox{and} \quad M_\KK\simeq \frac{M_s}{g_s^{1/4} \vo^{1/6}}\simeq \frac{M_p}{\vo^{2/3}} \,,\label{eq:basic}
\ee 
where $\vo$ is the Einstein-frame extra-dimensional volume in string units and perturbative reasoning assumes $g_s \ll 1$ and $\vo \gg 1$. 

\subsubsection*{Non-supersymmetric naturalness constraints}

For models without a suppression mechanism for the vacuum energy (like generic supersymmetric models) we have seen that technical naturalness keeps the inflationary scale from straying too far from the UV scale. 

For 4D models where the UV scale is $M_\KK$ we have: 
\be
M_\KK/\sqrt{4\pi} \lsim M_{\rm inf} \ll M_\KK\,,
\ee
with, as mentioned above, the first condition coming from asking  quantum corrections to be subdominant during inflation and the second bound from the validity of the EFT.  Therefore it is essentially impossible to simultaneosuly satisfy both conditions. At best we may still consider $M_{\rm inf} \sim M_{KK}$. Once $M_{\rm inf}$ is determined by inflationary observations we immediately learn a relation between $\cV$ and $g_s$:
\be 
\vo  \simeq 10^3 \left( \frac{0.01 M_p}{M_{\rm inf}} \right)^{3/2}\,,
\ee
where we take as benchmark $M_{\rm inf} \sim 10^{16} \, \hbox{GeV} \sim 0.01 \, M_p$,  corresponding to $H \sim 10^{14} \, \hbox{GeV} \sim 10^{-4} M_p$. Notice that for these numbers the conditions $g_s \ll 1$ and $\vo \gg 1$ do not allow too much room for $g_s$ and $\vo$.

Notice that this reasoning applies very generally, for any unwarped extra-dimensional model without fine-tuning and without a suppression mechanism for the vacuum energy. In particular it should be generic to non-supersymmetric constructions ({\em i.e.} with supersymmetry breaking for the 4D sector at or above the KK scale).

\subsubsection*{Supersymmetric naturalness conditions}

Consider next the situation where at least one supersymmetry breaks at a small enough scale to be described in the low-energy 4D theory, so that the 4D EFT is an $N=1$ 4D supergravity described by a K\"ahler potential, $K$, and superpotential, $W$. 

In particular we assume the gravitino mass $m_{3/2}(\Phi) \equiv e^{K/2} W/M_p^{2}$ is necessarily much smaller than $M_\KK$. (We write here explicitly the dependence on the various scalar moduli, $\Phi$, to emphasize we typically wish to work far from the potential minimum for inflationary applications.) It turns out, however that the condition the inflationary potential be below the UV scale, $V \ll M_\KK^4$, provides a stronger condition than does $m_{3/2} \ll M_\KK$.\footnote{See also \cite{Pramod} for similar considerations.} To see why recall that in 4D supergravity the potential can be written as:
\be
 V = |F(\Phi)|^{2}-3M_p^{2}m_{3/2}^{2}(\Phi)  \ll M_{\KK}^{4} \sim \frac{M_p^4}{\vo^{8/3}} \,, \label{eq:PotCons}
\ee
and in the absence of unnatural (functional) tunings the above bound is separately satisfied by each of the positive-definite terms in $V$. In particular, then, we see
\be
 m_{3/2}(\Phi) \ll \frac{M_\KK^2}{M_p} \sim \frac{M_p}{\vo^{4/3}} \ll M_\KK \,. \label{eq:m3/2bound}
\ee
On the other hand taking $M_{\rm inf}^4$ to be bigger than the generic size of quantum corrections, which in supersymmetric theories are of order $m_{3/2}^2 M^2 / (4\pi)^2$ with $M \sim M_\KK$, now implies $m_{3/2} M_\KK \ll 4\pi H M_p$, or the following lower limit on $H$:
\be
 M_{\rm inf}^4 \sim H^2 M_p^2 \gg \frac{m_{3/2}^2 M_\KK^2}{16\pi^2} \gg \frac{m_{3/2}^3 M_p}{16\pi^2}   \,,
\label{MIvsM32}
\ee
where the second inequality uses \pref{eq:m3/2bound}. This is easily satisfied\footnote{Actually the dimensional analysis changes if it happens that $H \gg m_{3/2}$ (as can be consistent with \pref{MIvsM32}). This is because with background curvature $R\sim H^{2}$ there is also a UV contribution of order $H^{2} M^{2}/16\pi^2$ , which can dominate. When it does the above bound instead degenerates to $M_\KK \ll 4\pi M_p$.} as long as the scale of supersymemtry breaking --- which is set by $m_{3/2}(\Phi)$ --- is far enough below the inflationary scale $M_{\rm inf}$.

\section{Generalised Fibre Inflation}
\label{sec:GFibInf}

This section asks a different kind of robustness question. Within the framework of Fibre Inflation models in  type IIB UV completion, this section explores the robustness of the construction, and how broadly the parameters for the low-energy inflationary potential can be varied. Although we find that inflation is robust, in so far as it occurs over a wider class of string constructions than in \cite{Fibre}, we find only marginal enhancement in the largest value ($r \lesssim 7 \times 10^{-3}$) found in \cite{Fibre}. This supports the robustness of this inferred upper limit for $r$ in these models. 

\subsection{Fibre Inflation revisited}

Fibre inflation \cite{Fibre} was discovered as a particular realisation of inflation in the general Large Volume Scenario (LVS) \cite{LVS} of moduli stabilisation of IIB CY orientifold compactifications. We here summarise the main properties of this scenario concentrating on the fibred CY case and inflationary applications.

The LVS scenario sits within type IIB string theory and exploits the well-developed tools \cite{GKP} that exist there for modulus stabilisation. The LVS focuses on weakly warped geometries and systematically organises the stabilisation of moduli order by order in $g_s$ and $\alpha'$: that is in powers of the string coupling and inverse powers of the volume of extra-dimensional cycles, $\tau_i$, in string units. Included in particular among these stabilised moduli is the total extra-dimensional volume $\vo$. 

One of the main results one finds in this program is that $\vo$ arises as the exponential of the volume of another cycle, $\tau_s$, where the validity of the $\alpha'$ expansion requires $\tau_s$ to be larger than unity (in string units), though it need not be enormously large. Ultimately $\tau_s$ is fixed by choices of flux quantum numbers and easily takes a range of moderately large values, and as a result the total volume $\cV$ samples an exponentially larger range. All other masses do so as well because they typically vary as a power of $\vo$. 

One is led to an interestingly varied hierarchy of masses, for which the most important dependence to track is usually the power of $\cV$. With 4D applications in mind it is useful to use 4D Planck units. As mentioned earlier the string scale then is $M_s \sim M_p/\sqrt\vo$ while the KK scale is $M_\KK \sim M_p/\vo^{2/3}$. By contrast the generic mass for the gravitino and moduli is much lighter, $m_{3/2} \sim w M_p/\cV$ and $M_{\rm mod} \sim M_p/\cV$, where $w$ is a dimensionless measure of the supersymmetry-breaking parameters appearing in the superpotential, $w \sim W/M_p^3$. 

\subsubsection*{A fibrous overview}

Before diving into the more detailed construction (and its generalizations), we first collect here the main points that motivate (and define) Fibre Inflation models within the LVS. The scenario seeks the inflaton among K\"ahler moduli because these moduli are the ones that only get stabilised by $\alpha'$ and $g_s$ effects, and (being moduli) should be light relative to the KK scale. When doing so two observations are central: ($i$) it is an `experimental' fact that most CY moduli correspond to fibrations (as defined in more detail below); and ($ii$) relative to other masses the fibration moduli first acquire their masses only at sub-dominant order in the $g_s$ and $\alpha'$ expansion. This has several important implications:
\begin{itemize}
\item The good news is that these moduli are systematically light, even relative to generic moduli, and so it is relatively easy to decouple all the other dangerous moduli from the inflationary dynamics. Closer inspection \cite{Fibre} shows the potential that generates their mass turns out to be of generic size $
 V \sim M_p^4/\vo^{10/3}$.

\item More good news is that the canonically normalised fields are often logarithms of the geometrical volumes of the corresponding cycles, making the potential depend exponentially on the canonical fields and ensuring their mass at the local minimum is of order $m_\phi \sim M_p/\vo^{5/3}$. This is also the generic order of magnitude of the Hubble scale, showing that $H$ scales with $\vo$ in the same way as the fibre-modulus mass $m_\phi$. But because the potential is exponential far from the minimum, $V \sim m_\phi^2 M_p^2 (1 - a e^{-b\phi/M_p} )$ with $a$ and $b$ $\mc{O}(1)$ constants, slow-roll is ultimately achieved along the lines forecast in \cite{ExpPot} because of the small size of $e^{-b\phi}$ rather than through any parametric hierarchy between $m_\phi$ and $H$. 

\item It is potentially bad news that in principle one needs to perform a string loop calculation to compute inflationary details. This is not quite as bad as it sounds, though, since the dependence on the variables of interest (such as $\vo$) can be inferred for the fibred geometry of interest starting from explicit calculations on toroidal spaces \cite{BHK} largely using scaling arguments \cite{BHP} and a proper matching with the low-energy Coleman-Weinberg potential \cite{Loop}. But there is also an upside: the attractive inflationary features only rely on a few robust features (the leading order kinetic term --- which determines the canonical variable and so leads to the potential's generic exponential form --- and the fact that the potential typically comes as inverse powers of the moduli). Furthermore typical uplifting terms in flux compactifications only depend on the overall modulus and dilaton but not on the other K\"ahler moduli giving rise naturally to a constant term in the scalar potential.
\end{itemize}

\subsubsection*{Model construction}

The total set of closed string moduli consists of the dilaton $S$, complex structure moduli $U$ and K\"ahler moduli $T$. The number of $U$ and $T$ fields changes with compactification but they are generically of the order of hundreds or thousands. Quantised fluxes of the two three-form fields present in IIB string compactifications generate a superpotential in the low-energy effective action that leads to the stabilisation of $S$ and all $U$ fields. The $T$ fields can be classified into at least two groups that can roughly be called `small' (or blow-up moduli) and `big' of which a good representative is a fibre modulus. It is known that most CY manifolds are fibrations of submanifolds (elliptic or K3 fibrations) \cite{GenericFibre}. A simple way to identify a fibre modulus is as follows: the overall volume of the manifold can be written as:
\be
\mc{V}=\frac 16 \,\kappa_{ijk} t_i t_j t_k\qquad i,j,k =1,\cdots h^{1,1}\,,
\label{v}
\ee
in which the $t_i$ are volumes of internal 2-cycles, $\kappa_{ijk}$ are the intersection numbers of these cycles and $h^{1,1}$ the corresponding Hodge number counting the number of 2 (and 4)-cycles. If a modulus $t_*$ appears only linearly in this expression then the corresponding manifold admits a K3 or a $T^4$ fibration over the base $\mbb{P}^1$ whose volume is given by $t_*$. If the Euler characteristic of the fibre is $\chi = 24$ then it is a K3 surface whereas if $\chi = 0$ the fibre is $T^4$ \cite{Math}. The volumes, $\tau_i$, of the 4-cycles dual to these 2-cycles are defined by $\tau_i = {\partial \vo}/{\partial t_i}$, 
These define the real part of the geometry's complex K\"ahler moduli:
\be
T_i = \tau_i + i \int_{D_i} C_4  \,, \,\, i=1,...,h^{1,1}  \,,
\label{Tmoduli}
\ee
where $D_i$ is the 4-cycle (divisor) whose volume is given by $\tau_i$ while $C_4$ is the Ramond-Ramond 4-form. The simplest realisation of a K3 fibration includes three K\"ahler moduli $t_1, t_2, t_3$ with:
\be
\vo =\lambda_1 t_1 t_2^2 + \lambda_2 t_3^3=\alpha\left(\sqrt{\tau_1}\tau_2 -\gamma \tau_3^{3/2}\right)= t_1\tau_1-\alpha\gamma\tau_3^{3/2},
\label{VolFib}
\ee
with $\alpha,\gamma$ simple functions of $\lambda_1,\lambda_2$, $\alpha = 1/ (2\sqrt{\lambda_1})$ and $\gamma = \frac23 \sqrt{\lambda_1 /(3\lambda_2)}$.\footnote{Recall that the K\"ahler cone condition of an exceptional two-cycle $t_3$ is $t_3<0$. This explains the negative sign in the second and third expression of $\vo$ in (\ref{VolFib}).} Topologically, this CY three-fold has a $\mbb{P}^1$ base of size $t_1$, a K3 or $T^4$ fibre of size $\tau_1$ and a point-like singularity resolved by a blow-up mode whose volume is given by $\tau_3$. For explicit CY three-folds with volume of the form (\ref{VolFib}) see \cite{CYmodels}.

For large volume models we restrict attention to orientifold projections that project out none of these K\"ahler moduli and focus on the large-volume regime, for which $t_1 \tau_1 \gg \alpha \gamma \tau_3^{3/2}$ in which case $ \vo \simeq t_1 \tau_1$.

The scalar potential is determined by the expressions for the K\"ahler and superpotential:
\be
K=-2\ln\left(\vo +\frac{\zeta}{2g_s^{3/2}}\right) \qquad W=W_0 + A\,e^{-aT_3}\,,
\label{K}
\ee
with $W_0$, $A$ and $g_s$ determined by the fluxes after the stabilisation of $S$ and the $U$ fields. Here $\zeta$ and $a$ are model dependent constants. Notice that the fields $T_1$ and $T_2$ only appear in the combination $\vo$. This immediately implies that at this stage of approximation (leading order in perturbation theory) one combination of $T_1$ and $T_2$ remains flat. This is the candidate for an inflaton. The scalar potential after stabilising $S$ and the $U$ fields and the axionic components of the $T_3$ field, looks like:
\be\label{V0up}
V=8a^2A^2\frac{\sqrt{\tau_3}}{3\alpha\gamma\vo} e^{-2a\tau_3}-4aAW_0\frac{\tau_3}{\vo} e^{-a\tau_3} + \frac{3\zeta W_0^2}{4g_s^{3/2}\vo^3}+ V_{\rm up}\,,
\ee
where the phase of $W_0$ is absorbed in the stabilisation of the imaginary part of $T_3$. $V_{\rm up}$ is the uplift term in the scalar potential. Several sources of $V_{\rm up}$ have been identified ranging from anti D3 branes \cite{KKLT} (for recent developments see \cite{new}), T-branes \cite{Tbranes}, non-perturbative effects on hidden D3s \cite{NPD3s} etc. For our purposes we will only use it to enable the tuning of the final minimum of the scalar potential after adding the string loop corrections discussed later, to essentially zero. At the minimum the volume $\vo$ and $\tau_3$ have the standard large volume values:
\be
\vo\sim W_0 \,e^{a\tau_3} \qquad \text{and }\qquad \tau_3\sim g_s^{-1}\,.
\ee
As mentioned one combination of $\tau_1$ and $\tau_2$ is not determined at this level of approximation. This remaining flat direction can be lifted by including subleading string loop corrections to the K\"ahler potential \cite{Fibre, GeneralLVS}. Let us analyse the structure of these corrections in order to evaluate the robustness of this inflationary model. 

\subsubsection*{String loops}

The leading string loop effects arise at order $\mc{O}(\alpha'^2 g_s^2)$, and so they are both $g_s$ and $\alpha'$ corrections to the effective action. They have been computed explicitly only for simple toroidal orientifold cases like $N=1$ compactifications on $T^6/(\mathbb{Z}_2\times\mathbb{Z}_2)$ where they take the form \cite{BHK}:
\be
K_{g_s} = K_{\rm 1-loop}^{\KK} + K_{\rm 1-loop}^{\ssW}\,.
\label{Kgs}
\ee
The two contributions in (\ref{Kgs}) have a different microscopic origin since $K_{\rm 1-loop}^{\KK}$ originates from a 1-loop diagram of open strings stretched between D7-branes (or O7-planes) and D3-branes (or between non-intersecting D7-branes), and in the closed-string channel can be interpreted as due to the tree-level exchange of closed strings carrying KK momentum. On the other hand,  $K_{\rm 1-loop}^{\ssW}$ comes from the exchange of closed strings wound around a non-contractible 1-cycle at the intersection between different stacks of D7-branes (or between D7-branes and O7-planes). In 4D Einstein frame they look like (with $s\equiv {{\rm Re}(S)}$):
\be
K_{\rm 1-loop}^{\KK}= \sum_{i=1}^3 \frac{\mc{C}_i^{\KK}(U)}{s \,\tau_i} \qquad\text{and}\qquad 
K_{\rm 1-loop}^{\ssW} = \sum_{i\neq j =1}^3\frac{\mc{C}_{ij}^{\ssW}(U)}{\tau_i\tau_j}\,,
\label{K1loop}
\ee
where $\mc{C}_i^{\KK}(U)$ and $\mc{C}_{ij}^{\ssW}(U)$ are complicated functions of the complex structure moduli which involve Eisenstein series. Notice that both $K_{\rm 1-loop}^{\KK}$ and $K_{\rm 1-loop}^{\ssW}$ correctly scale as $g_s^2$ in string frame since $\langle s \rangle = g_s^{-1}$ and $\tau_{\tiny{\rm str}} = g_s \tau$. Thus in the original string frame these corrections scale as (fixing the dilaton and considering all 4-cycles of the same size): 
\be
K_{\rm 1-loop}^{\KK} \sim  \frac{g_s^2}{\tau_{\tiny{\rm str}}}\,, \qquad K_{\rm 1-loop}^{\ssW} \sim \frac{g_s^2}{\tau_{\tiny{\rm str}}^2}\,,
\qquad\Rightarrow \qquad \frac{K_{\rm 1-loop}^{\KK}}{K_{\rm 1-loop}^{\ssW}}\sim \tau_{\tiny{\rm str}} \gg 1\,,
\label{Kscaling}
\ee
implying that in the regime where the EFT is under control KK corrections are dominant with respect to the winding ones. Moreover, for an arbitrary CY compactification, $K_{\rm 1-loop}^{\KK}$ is more generic than $K_{\rm 1-loop}^{\ssW}$ since KK states are a ubiquitous feature of string compactifications whereas the presence of intersecting stacks of branes and non-contractible 1-cycles at their intersection locus are features which depend both on the particular brane configuration and on the topology of the internal space.

Let us stress also that the volume scaling of $K_{\rm 1-loop}^{\KK}$ can be estimated via a simple low-energy argument \cite{Loop}. String loop effects should reproduce standard QFT loop corrections at low energies. These generate corrections to the scalar kinetic terms which are suppressed by the coupling of the gauge interaction these scalars couple to. For gauge theories living on D7-branes the corresponding gauge coupling is given by the 4-cycle $\tau$ wrapped by the D7-brane whereas for D3-branes the gauge coupling is given by the dilaton. Given that the kinetic terms are derived by taking second derivatives of the K\"ahler potential, we have:
\be
\frac{\partial^2\left(K_{\rm 1-loop}^{\KK}\right) }{\partial \tau^2} \sim \frac{g_{\tiny{D7}}^2}{16\pi^2} \frac{\partial^2\left(K_{\rm tree}\right) }{\partial \tau^2}\qquad\text{with}\qquad g_{\tiny{D7}}^{-2} = \tau\,,
\label{KD7}
\ee
and:
\be
\frac{\partial^2\left(K_{\rm 1-loop}^{\KK}\right) }{\partial s^2} \sim \frac{g_{\tiny{D3}}^2}{16\pi^2} \frac{\partial^2\left(K_{\rm tree}\right) }{\partial s^2}\qquad\text{with}\qquad g_{\tiny{D3}}^{-2}= s\,,
\label{KD3}
\ee
which imply:
\be
K_{\rm 1-loop}^{\KK} \sim \frac{1}{16\pi^2 s \tau}\,.
\label{KKKest}
\ee
Clearly (\ref{KKKest}) reproduces the exact $s$ and $\tau$-dependence of $K_{\rm 1-loop}^{\KK}$ in (\ref{K1loop}). However this simple logic cannot be followed to estimate the volume scaling of $K_{\rm 1-loop}^{\ssW}$ since at low energy we do not expect to see the effect of corrections due to the exchange of winding strings as $m_\ssW > M_s > m_\KK$.

Rewriting the scaling relations (\ref{Kscaling}) as:
\be
K_{\rm 1-loop}^{\KK} \sim  \frac{g_s^2\,t_{\tiny{\rm str}}}{\vo_{\tiny{\rm str}}}\,, \qquad\text{and}\qquad 
K_{\rm 1-loop}^{\ssW} \sim \frac{g_s^2}{t_{\tiny{\rm str}}\,\vo_{\tiny{\rm str}}}\,,
\label{Kscaling2}
\ee
and noticing that the KK and winding mass scales can be written respectively as $m_\KK^2\sim t_{\tiny{\rm str}}^{-1}$ and $m_\ssW^2\sim t_{\tiny{\rm str}}$, the toroidal results (\ref{K1loop}) can be rewritten in Einstein frame as:
\be
K_{\rm 1-loop}^{\KK}= \sum_i \frac{\mc{C}_i^{\KK}(U)\, m_{\KK,i}^{-2}}{s\, \vo} \qquad\text{and}\qquad 
K_{\rm 1-loop}^{\ssW} = \sum_{i\neq j}\frac{\mc{C}_{ij}^{\ssW}(U)\, m_{\ssW,ij}^{-2}}{\vo}\,,
\label{K1loopCY}
\ee
where $m_{\KK,i}^{-2}\sim t_i$ is the 2-cycle transverse to the D7-brane wrapped around $\tau_i$ while $m_{\ssW,ij}^{-2}\sim \tau_i\cap \tau_j \sim t_{ij}$ is the 2-cycle where the two D7-branes wrapped around $\tau_i$ and $\tau_j$ intersect. The expressions (\ref{K1loopCY}) reflect now clearly the understanding of these effects as due to the tree-level exchange of KK and winding strings (the $\vo$-factor comes from the Weyl rescaling to 4D Einstein frame). Moreover \cite{BHP} used the results (\ref{K1loopCY}) to conjecture the form of the string 1-loop corrections to the K\"ahler potential for an arbitrary CY compactification where now the index $i$ runs from $1$ to the total number of wrapped D7-branes while $k$ goes from $1$ to the total number of intersections between stacks of D7-branes. This logic does not allow us to determine the exact moduli-dependence of $\mc{C}_i^{\KK}(U)$ and $\mc{C}_k^{\ssW}(U)$ but this is not a problem since the complex structure moduli are stabilised at tree-level by background fluxes, and so these two unknown functions can just be regarded as $\mc{O}(1)$  constants. 

The conjectured expressions (\ref{K1loopCY}) suggest that string 1-loop corrections to the K\"ahler potential for an arbitrary CY are homogeneous functions of the 2-cycle moduli of degree $n$ ($n=-2$ for $K_{\rm 1-loop}^{\KK}$ while $n=-4$ for $K_{\rm 1-loop}^{\ssW}$). Using this piece of information, \cite{Loop} showed that the leading order contribution of each of these effects to the scalar potential behaves as: 
\be
V_{{\rm loop}} = -\frac{W_0^2}{\vo^2}\frac{n}{4} \left(n+2\right) K_{{\rm 1-loop}} + \cdots \,.
\label{Vloop}
\ee
We immediately realise that in the KK case with $n=-2$ there is a leading order cancellation which \cite{Loop} dubbed \textit{extended no-scale structure}. This cancellation does not take place for winding corrections which could be naively considered as the leading order effect in the scalar potential. However, as we showed in (\ref{Kscaling}), $K_{\rm 1-loop}^{\KK}$ dominates over $K_{\rm 1-loop}^{\ssW}$ for large cycle volumes, and so we need to take into account also the first non-vanishing KK contribution to $V$. Because of the extended no-scale cancellation, this can originate from both subleading 1-loop contributions and leading 2-loop effects. The subleading 1-loop contribution to $V$ has been derived in \cite{Loop} and reads:
\be
V_{{\rm 1-loop}}^{\KK} = \frac{W_0^2}{\vo^2}\sum_{i,j}\frac{\mc{C}_i^{\KK}}{s}\frac{\mc{C}_j^{\KK}}{s} K_{{\rm tree},ij} + \cdots \,.
\label{V1KK}
\ee
Noticing that $K_{\rm 1-loop}^{\KK}$ in (\ref{K1loopCY}) can be rewritten as:
\be
K_{\rm 1-loop}^{\KK}= \sum_i \frac{\mc{C}_i^{\KK}(U)\, t_i}{s\, \vo} = - \sum_i \frac{\mc{C}_i^{\KK}}{s} K_{{\rm tree},i}\,,
\label{K1loopCY2}
\ee
we see that the total 1-loop contribution to the scalar potential can be written as an expansion in derivatives of the tree-level K\"ahler metric as:
\be
V_{{\rm 1-loop}}^{\KK} = \frac{W_0^2}{\vo^2}\left[\alpha_1 \sum_i \frac{\mc{C}_i^{\KK}}{s} K_{{\rm tree},i}+\alpha_2\sum_{i,j}\frac{\mc{C}_i^{\KK}}{s}\frac{\mc{C}_j^{\KK}}{s} K_{{\rm tree},ij} + \alpha_3\sum_{i,j,k}\frac{\mc{C}_i^{\KK}}{s}\frac{\mc{C}_j^{\KK}}{s}\frac{\mc{C}_k^{\KK}}{s} K_{{\rm tree},ijk}+\cdots \right] \nn
\label{Vexp}
\ee
where $\alpha_1 = \frac{n}{4} \left(n+2\right) = 0$, $\alpha_2=1$ and $\alpha_3$ an $\mc{O}(1)$ constant. As shown in \cite{Loop} for different CY examples, the terms in the expansion (\ref{Vexp}) match the volume scaling of the terms of the low energy 1-loop Coleman-Weinberg potential:
\be
V_{{\rm 1-loop}}^{{\scriptscriptstyle CW}} = \frac{1}{64\pi^2}\left[\Lambda^4 {\rm Str}\left(M^0\right)\ln\left(\frac{\Lambda^2}{\mu^2}\right) + 2\Lambda^2{\rm Str}\left(M^2\right) +  {\rm Str}\left(M^4\ln\left(\frac{\Lambda^2}{M^2}\right)\right)\right]\,,
\label{VCW}
\ee
when the cut-off $\Lambda = m_\KK$ and ${\rm Str}\left(M^2\right)\simeq m_{3/2}^2$ are written in terms of the K\"ahler moduli. Moreover, the first term in (\ref{VCW}) has a vanishing coefficient since ${\rm Str}\left(M^0\right) = 0$ in any supersymmetric theory (even if SUSY is broken), so providing a better understanding of the extended no-scale cancellation based on supersymmetry. 

Due to this leading order cancellation of the 1-loop KK contribution to the scalar potential, 2-loop corrections could also give rise to competing effects. In fact, (\ref{V1KK}) scales as:
\be
V_{{\rm 1-loop}}^{\KK} \sim  \frac{W_0^2}{\vo^2}\,\delta K \qquad\text{where}\qquad \delta K \sim \frac{K_{\rm 1-loop}^{\KK}}{s\,\tau}\,,
\ee 
which in string frame and for fixed dilaton gives a term which behaves as a 2-loop correction since:
\be
\delta K \sim \frac{g_s^2}{\tau_{\tiny{\rm str}}}\,K_{\rm 1-loop}^{\KK}\sim K_{\rm 2-loop}^{\KK}\,.
\label{K2}
\ee
The volume scaling of 2-loop KK corrections to the K\"ahler potential used in (\ref{K2}) can be estimated by following the same logic used in (\ref{KD7}) and (\ref{KD3}):
\be
\frac{\partial^2\left(K_{\rm 2-loop}^{\KK}\right) }{\partial \tau^2} \sim \frac{g_{\tiny{D7}}^2}{16\pi^2} \frac{\partial^2\left(K_{\rm 1-loop}^{\KK}\right) }{\partial \tau^2}\,,
\label{KD72}
\ee
and:
\be
\frac{\partial^2\left(K_{\rm 2-loop}^{\KK}\right) }{\partial s^2} \sim \frac{g_{\tiny{D3}}^2}{16\pi^2} \frac{\partial^2\left(K_{\rm 1-loop}^{\KK}\right) }{\partial s^2}\,,
\label{KD32}
\ee
which imply a $g_s$ and volume scaling in perfect agreement with (\ref{K2}):
\be
K_{\rm 2-loop}^{\KK} \sim \frac{1}{16\pi^2 s^2 \tau^2} \sim \frac{K_{\rm 1-loop}^{\KK}}{16\pi^2 s \tau}\,.
\label{KKKest2}
\ee
It is therefore sensible to expect that 2-loop KK corrections at linear order behave as 1-loop KK corrections at quadratic order even if there is no exact toroidal computation at 2-loop level which we could try to generalise to the arbitrary CY case.

\subsubsection*{Inflationary potential}

Applying these considerations to our K3 or $T^4$-fibred case, we find that 1-loop winding corrections to $V$ read:
\be
V_{{\rm 1-loop}}^{\ssW} = -2\,\frac{W_0^2}{\vo^2}\,K_{{\rm 1-loop}}^{\ssW} +\cdots
= -\,\frac{W_0^2}{\vo^2}\,\frac{B}{\vo\sqrt{\tau_1}}+\cdots\qquad\text{with}\qquad B = 4\alpha\mc{C}_{12}^{\ssW}\,,
\label{V1w}
\ee
while the combined effect of 1- and 2-loop KK corrections looks like:
\be
V_{{\rm 1-loop}}^{\KK} + V_{{\rm 2-loop}}^{\KK} = g_s^2\,\frac{W_0^2}{\vo^2} \left(\frac{A}{\tau_1^2}+\frac{C\,\tau_1}{\vo^2}\right)+\cdots\,,
\label{Vgs1}
\ee
where (calling the coefficients of the 2-loop effects as $\mc{D}_i^\KK$): 
\be
A = \left(\mc{C}_1^\KK\right)^2 +\mc{D}_1^\KK \qquad\text{and}\qquad C = 2 \alpha^2\left(\mc{C}_1^\KK\right)^2 + \mc{D}_2^\KK\,.
\ee
Notice that we cannot determine the sign of the three coefficients $A$, $B$ and $C$ which we however expect to be $\mc{O}(1)$ numbers. Parameterising the flat direction to be lifted as $\tau_1$, the minimum of the total string loop potential lies at:
\be
\langle\tau_1\rangle^{3/2} = \left(\frac{8 g_s^2 A \vo}{B}\right)\left(1+\frac{B}{|B|}\sqrt{1+ 32 g_s^4\,\frac{A C}{B^2}}\right)^{-1}\,.
\ee
For $g_s^4\ll 1$ we have:
\be
\langle\tau_1\rangle^{3/2} \simeq g_s^2\left(\frac{4 A}{B}\right) \vo\quad\text{for}\,B>0\qquad\text{and}\qquad
\langle\tau_1\rangle^{3/2} \simeq g_s^{-2}\left(\frac{|B|}{2 C}\right) \vo\quad\text{for}\,B<0\,, \nn
\ee
which require $A>0$ for $B>0$ and $C>0$ for $B<0$. Notice that these conditions are always satisfied if in (\ref{Vgs1}) the first non-vanishing 1-loop KK contribution dominates over the 2-loop effect. Rewriting these minima in terms of the original fields $\tau_1$ and $\tau_2$ we have:
\begin{itemize}
\item $B>0$:
\be
\langle\tau_1\rangle \simeq g_s^2\left(\frac{4\alpha A}{B}\right) \langle\tau_2\rangle \ll \langle\tau_2\rangle \qquad\text{for}\quad g_s\ll 1
\label{minimaPos}
\ee

\item $B<0$
\be
\langle\tau_1\rangle \simeq g_s^{-2}\left(\frac{|B|}{2 C}\right) \langle\tau_2\rangle \gg \langle\tau_2\rangle\qquad \text{for}\quad g_s\ll 1\,.
\label{minimaNeg}
\ee
\end{itemize}
Therefore in the case with $B>0$ inflation should take place from right to left with the inflaton $\tau_1$ that during inflation relaxes from larger to smaller values, while in the case with $B>0$ during inflation $\tau_1$ increases from smaller to larger values. Keeping the volume fixed, canonical normalisation gives \cite{Loop}:
\be
\tau_1 = e^{k\phi}\qquad\text{with}\qquad k =\frac{2}{\sqrt{3}}\,.
\ee
Substituting this relation in the string loop potential and expanding the inflaton around its minimum as $\phi=\langle\phi\rangle+\hat\phi$, we have (adding also the positive uplifted term (\ref{V0up}) which is then tuned to get a zero cosmological constant at the minimum):
\begin{itemize}
\item $B>0$:
\be
V=\frac{V_0}{\vo^{10/3}} \left[3-  4  \,e^{-k\hat\phi/2}+e^{-2 k\hat\phi} +R \left(e^{k\hat\phi}-1\right)\right]\,,
\label{VBpos}
\ee
where:
\be
V_0 = W_0^2\,\left(\frac{|B|^4}{256 g_s^2 |A|}\right)^{1/3}\qquad\text{and}\qquad R = 16 g_s^4 \frac{|A| C}{|B|^2}\ll 1
\ee

\item $B<0$:
\be
V=\frac{V_0}{\vo^{10/3}} \left[2  \,e^{-k\hat\phi/2} -3 + e^{k\hat\phi}+R\left(e^{-2 k\hat\phi}-1\right)\right]\,,
\label{VBneg}
\ee
where:
\be
V_0 = W_0^2\,\left(\frac{g_s^2 |C| |B|^2}{4}\right)^{1/3}\qquad\text{and}\qquad R = 4 g_s^4 \frac{A C}{|B|^2}\ll 1
\ee
\end{itemize}

\begin{figure}[!ht]
\centering
\includegraphics[height=60mm,width=70mm]{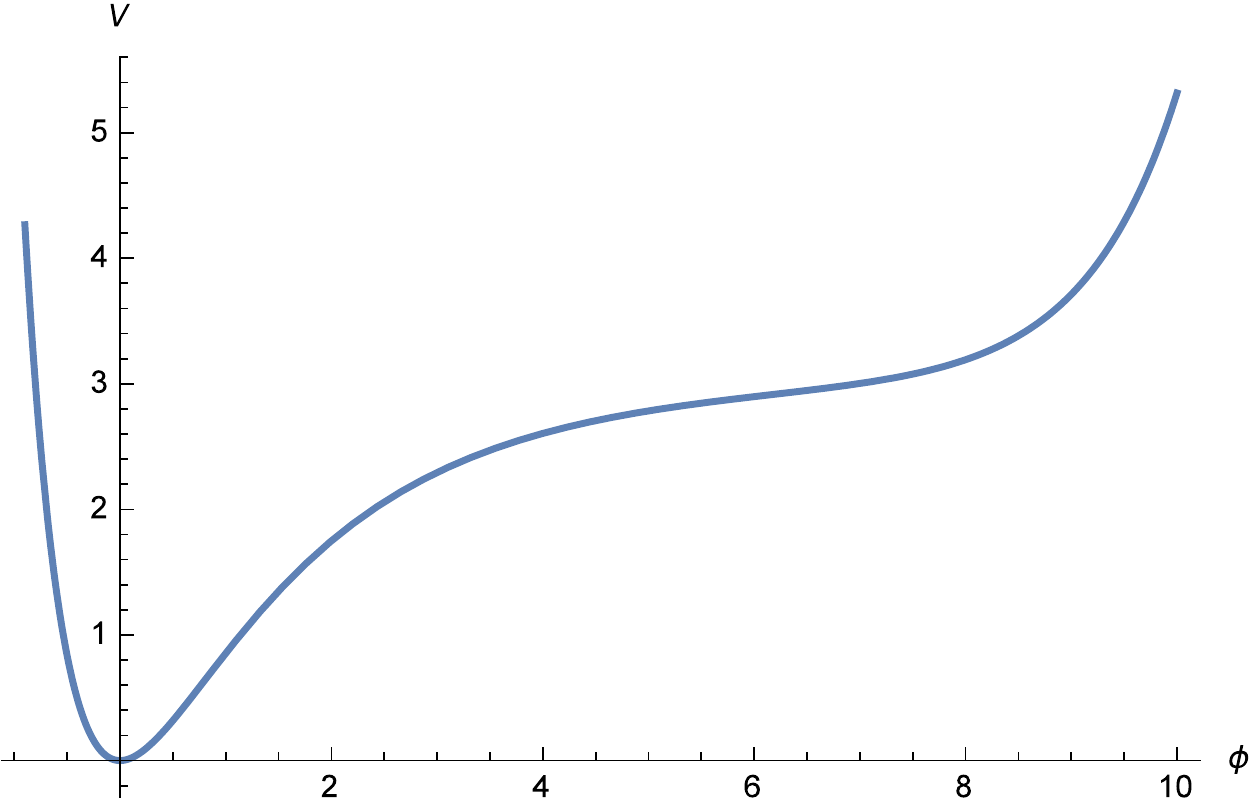}
\label{Fig1}
\caption{$V$ versus $\hat\phi$ for $k=2/\sqrt{3}$ and $R=2.25\cdot 10^{-5}$.} 
\end{figure}

\begin{figure}[!ht]
\centering
\includegraphics[height=60mm,width=70mm]{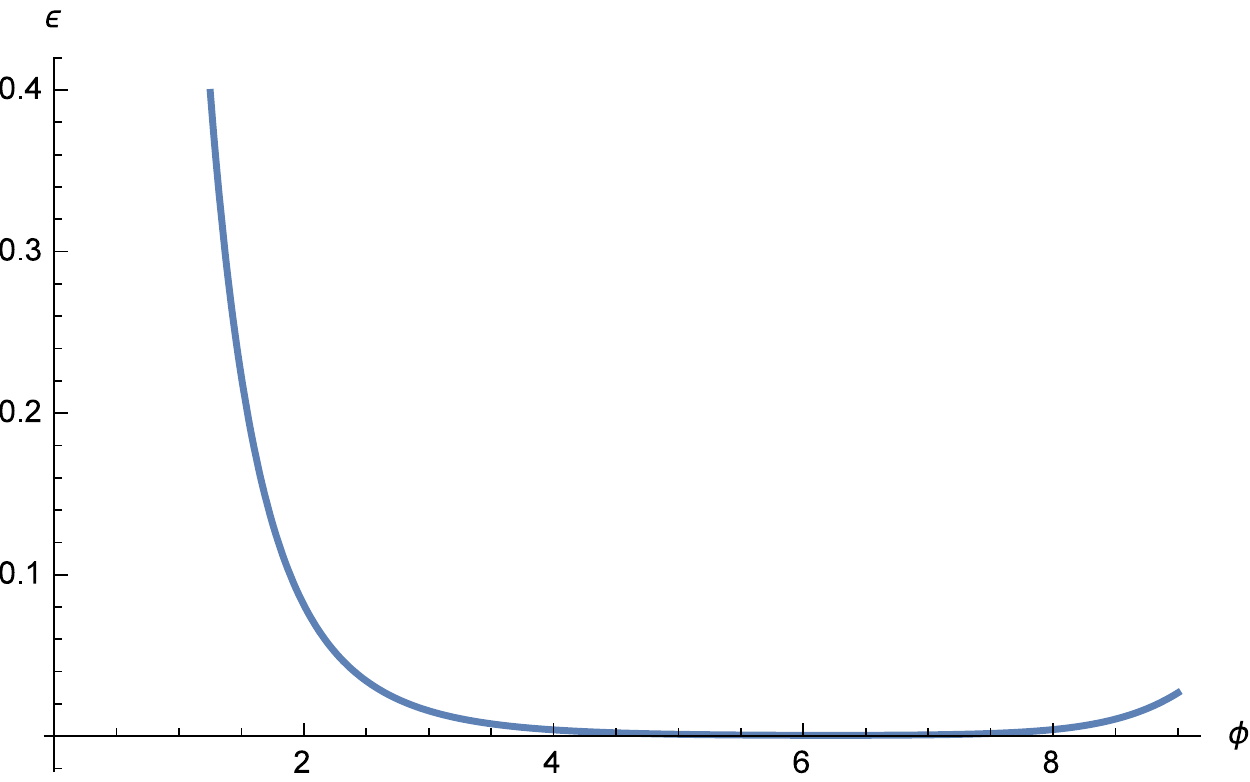}
\includegraphics[height=60mm,width=70mm]{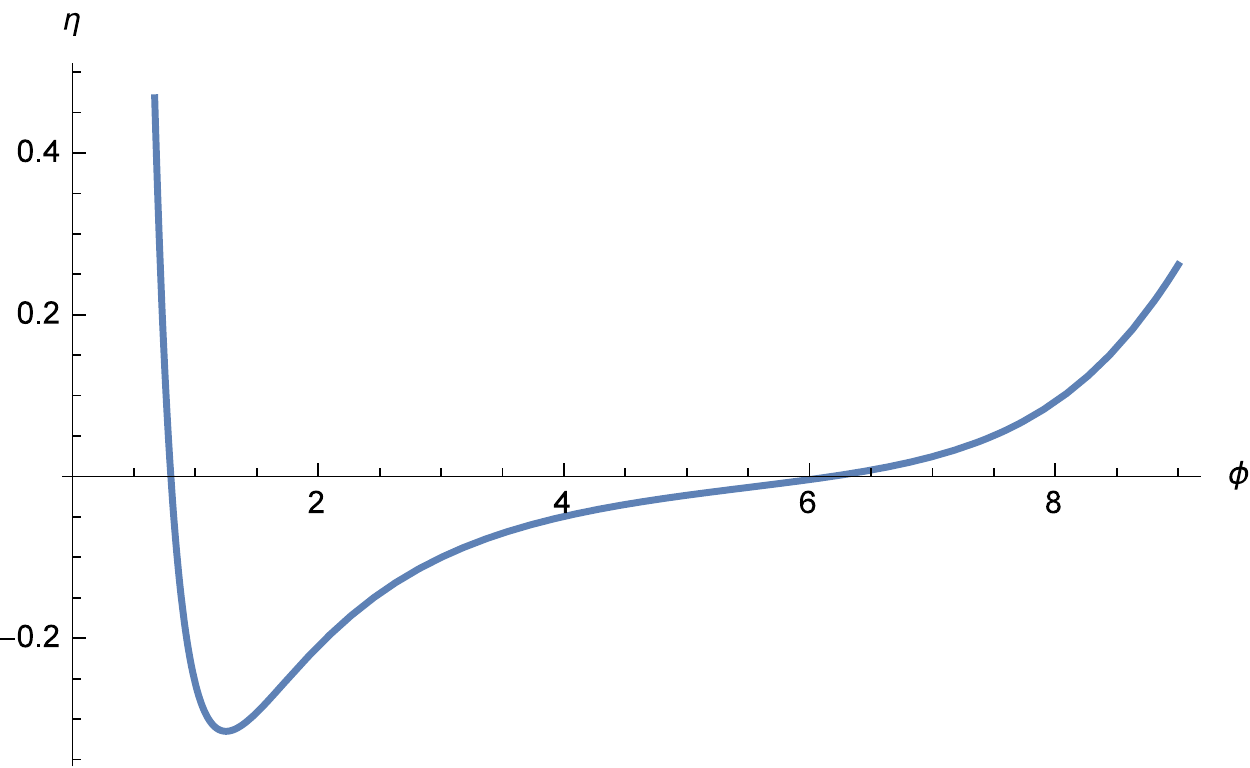}
\label{Fig2}
\caption{$\epsilon$ and $\eta$ versus $\hat\phi$ for $k=2/\sqrt{3}$ and $R=2.25\cdot 10^{-5}$.} 
\end{figure}

Both of the potentials (\ref{VBpos}) and (\ref{VBneg}) are flat enough to drive inflation. In the case with $B>0$ inflation takes place for positive values of $\hat\phi$ and during inflation the original fibre modulus $\tau_1$ decreases in size. On the other hand for $B<0$ $\hat\phi$ is negative during inflation and $\tau_1$ increases in size. The two potentials give rise to a different inflationary phenomenology. Given that $R$ is naturally small in the region where perturbation theory is under control, i.e. for $g_s\ll 1$, for $B>0$ the potential (\ref{VBpos}) features a plateau region at large $\hat\phi$ where the inflationary potential can be approximated as:
\be
V=\frac{V_0}{\vo^{10/3}} \left(3-  4  \,e^{-k\hat\phi/2} \right)\,.
\label{VBposApp}
\ee
This gives the following simple relation between $r$ and $n_s$:
\be
r \simeq \frac{8}{k^2} \left(n_s-1\right)^2 = 6 \left(n_s-1\right)^2\,,
\label{rns}
\ee
which is a particular example of the general relation (\ref{rvsns}) for an effective `decay constant' $f = \sqrt{3} \,M_p$. The total potential (\ref{VBpos}) with $R=2.25\cdot 10^{-5}$ is plotted in Fig. \ref{Fig1} while Fig. \ref{Fig2} gives the behaviour of $\epsilon$ and $\eta$.

On the other hand for $B<0$ the potential (\ref{VBneg}) in the inflationary regime for negative $\hat\phi$ behaves as a negative exponential:
\be
V=\frac{2\,V_0}{\vo^{10/3}}\,e^{-k\hat\phi/2} \,,
\label{VBnegApp}
\ee
which leads to a clear prediction for $r$ that is in disagreement with observations \cite{planck15} since:
\be
\epsilon \simeq \frac{k^2}{8} = \frac 16\qquad\Rightarrow\qquad  r = \frac 83\,.
\ee
The case with $B<0$ is therefore experimentally ruled out.

\subsection{Robustness}

Let us make a few comments on the robustness of these models:
\begin{itemize}
\item KK loop corrections are generically present in any CY compactification while winding loop corrections are more model dependent since they depend on the brane setup and the topology of the internal space.

\item Winding loop corrections are under better control than KK loops since, due to the extended no-scale cancellation, 1- and 2-loop KK effects lead to competing contributions to the scalar potential.

\item When they are present and have the correct sign, winding loop corrections generate a plateau region which is suitable to drive inflation. The robust prediction of this inflationary scenario is the relation (\ref{rns}) between $r$ and $n_s$.

\item In order to have a model, instead of just a scenario, with an exact prediction for $r$ and $n_s$, we need to add KK loop corrections which develop a minimum of the inflationary potential. 

\item Even if KK loops are under less control than winding effects, the total inflationary potential (\ref{VBpos}) is still robust since: 
\begin{enumerate}
\item String loops are both $\vo$ and $g_s$-suppressed with respect to $\alpha'$ effects which develop a potential for the volume mode. This ensures the presence of a mass hierarchy between the inflaton $\tau_1$ and all the other moduli. The inflaton is at leading order a flat direction, and so it is flat enough to drive inflation and all the other moduli can be safely decoupled from the inflationary dynamics. Moreover, there are no problems with trans-Planckian values of $\hat\phi$ since higher order operators are suppressed due to an approximate shift symmetry for $\hat\phi$ (broken only by small loop effects) \cite{NCNInf}.

\item Perturbation theory is under control throughout all the inflationary dynamics and also in the minimum since $g_s\ll 1$ and both $\tau_1$ and $\tau_2$ are always in the large volume regime. Thus higher order winding and KK loops are subdominant with respect to the leading effects which generate the potential (\ref{VBpos}). In particular the requirement to match the observed amplitude of the density perturbations fixes $\vo \lesssim 10^4$ for $g_s \simeq 0.1$. This value of the internal volume in turn sets all the relevant energy scales: the string scale $M_s  \sim 5 \cdot 10^{15}$ GeV, the KK scale $M_\KK \sim 10^{15}$ GeV, the inflationary scale $M_{\rm inf} \sim V_{\rm inf}^{1/4} \sim 10^{14}$ GeV and the Hubble scale $H_{\rm inf} \sim V_{\rm inf}^{1/2}/M_p \sim 10^{10}$ GeV. Thus the EFT is under control. The separation in energy between $M_{\rm inf}$ and $M_\KK$ becomes smaller for cases with larger values of $r$ where the EFT is therefore only marginally under control. 
\end{enumerate}

\item In the CY cases where winding corrections are absent, an inflationary potential with a structure similar to the one in (\ref{VBposApp}) can be obtained by including higher derivative $\alpha'$ effects which are generically present in any CY compactification \cite{F4}. In fact, \cite{Broy:2015zba} used these $F^4$ terms to generate potentials with an inflationary plateau which can lead to:
\be
r \simeq \frac 32 \left(n_s-1\right)^2\qquad\text{or}\qquad r \simeq 6 \left(n_s-1\right)^2\,,
\label{rns2}
\ee
corresponding respectively to effective `decay constants' $f=\sqrt{3} \,M_p/2$ and $f = \sqrt{3} \,M_p$. In both cases,  similarly to \cite{Loop}, the minimum of the total inflationary potential is obtained by including KK loop effects. We therefore conclude that the general relation (\ref{rvsns}) is a robust prediction of this class of models since it holds also in the absence of winding loop corrections. The exact values of the effective `decay constant' $f$ and the spectral index $n_s$ are instead more model-dependent features which depend on the particular effects used to develop the inflationary potential and the topology of the underlying CY space. Another example of an inflationary model satisfying the $r$-$n_s$ relation (\ref{rvsns}) but with a larger effective `decay constant' $f$ is presented in Appendix \ref{AppB}. 
\end{itemize}

\subsection{Comparison with other models}

\begin{figure}[!ht]
\centering
\includegraphics[height=80mm,width=100mm]{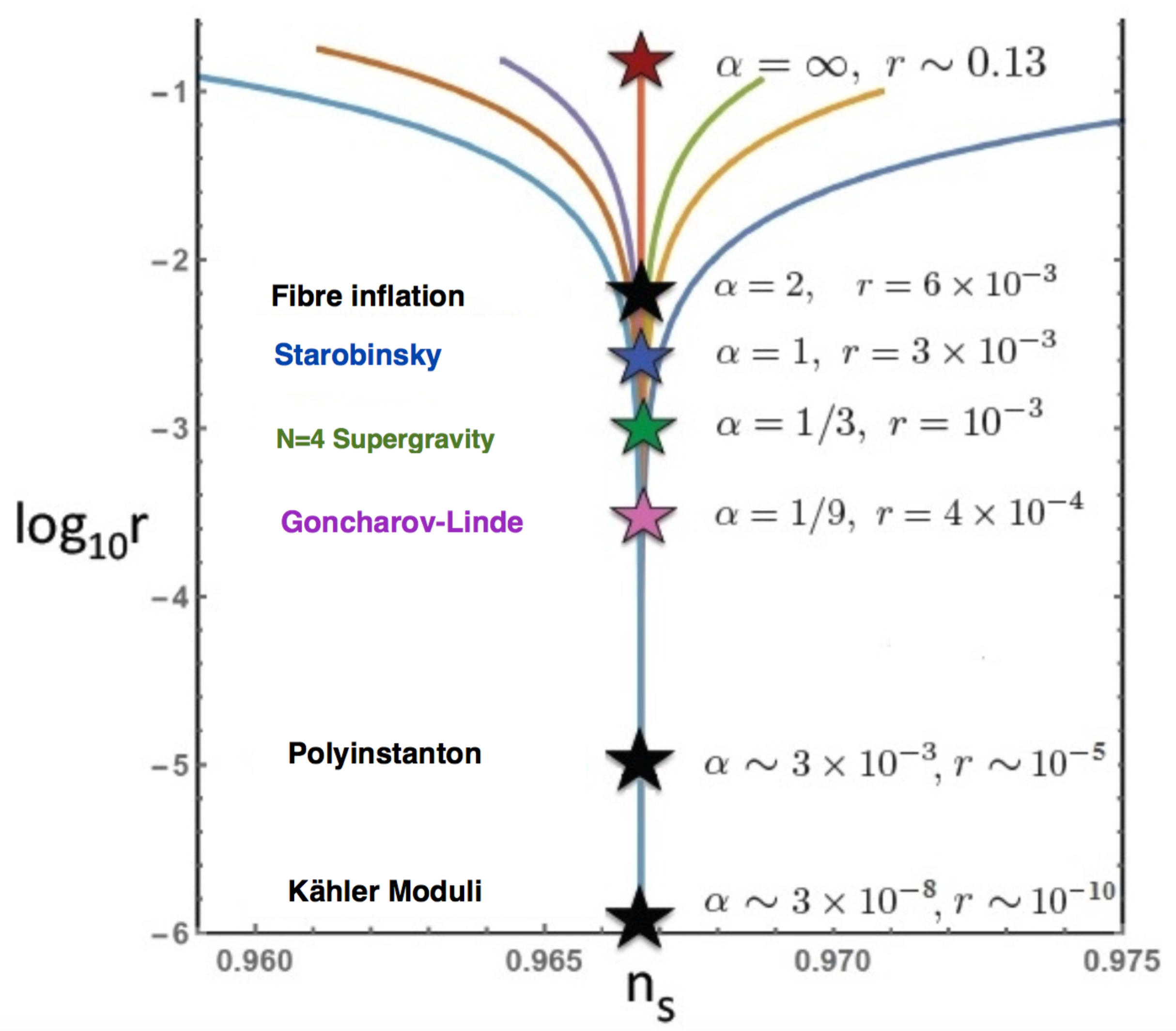}
\label{Fig3}
\caption{Fibre inflation fits in the $\alpha$ attractors class of models corresponding to $\alpha=2$ and can be seen as a stringy realisation of $\alpha$ attractors (figure adapted from \cite{ckl}). Other string scenarios such as K\"ahler moduli inflation and Poly-instanton inflation are also in this class but with much smaller values of $r$ and then unobservable tensor modes.}
\end{figure}

Fibre Inflation, as well as the original K\"ahler \cite{kahlerinflation}\, and polyinstanton \cite{polyinstanton}\, inflation scenarios, can be seen as stringy realisations of a general class of potentials of the form $V \sim V_0 - V_1 \, e^{-(\phi/f)^n}+\cdots$; and (as argued earlier) because this class includes exponential potentials it also includes the Starobinsky model and what have come to be called `$\alpha$-attractors'. Models in this class with $n = 1$ have $V' \sim V''$ and so $\epsilon \sim \eta^2$ which gives the $r$-$n_s$ relation:
\be
r= 3\alpha \left(n_s-1\right)^2\,,
\ee
with $\alpha$ given as in the introduction in terms of $f/M_p$. Models with larger $n$ do not quite so simply relate $V'$ and $V''$, but usually predict smaller values for $r$. Fig.~\ref{Fig3} --- adapted from \cite{ckl} --- plots these predictions in the $r$-$n_s$ plane, showing the range of $r$ that would have to be probed to distinguish several benchmark models. 

Stringy realisations identified to date only cover a relatively small range of values of $n_s$ and $r$ in this plot, since arbitrary values of $\alpha$ are not (yet) available from string constructions. So far two general classes of stringy models have emerged: in one $r$ is essentially zero and never within observational reach; for the other --- which includes Fibre Inflation \cite{Fibre} (for which $\alpha=2$ and $r$ is predicted to be $r\sim 6\times 10^{-3}$) --- fits precisely in the observationally interesting region that will be testable in the next few years. Observation of tensor modes (and the measurement of $r$) could therefore provide a good way to distinguish amongst the various proposals. 

\subsection{Generalising Fibre Inflation}

In this subsection we will investigate how to generalise the Fibre Inflation model. The general idea is to look for inflatons as real parts of K\"ahler moduli, as in K\"ahler and Fibre Inflation, in as model independent a way as possible. The point we have emphasised before is that, focusing on the tree-level and including also the leading $\alpha'$ correction (proportional to quartic terms in the internal curvature), the K\"ahler potential depends only on one combination of the many moduli, that is the overall volume $\vo$. Thus, at this level of approximation, the scalar potential depends just on $\vo$, and the leading order term in $V$, after the volume is stabilised, gives a constant. The dependence of $V$ on the other K\"ahler moduli comes from subleading effects like string loop corrections \cite{BHK, Loop, GeneralLVS}, D-terms \cite{FI}, $F^4$ terms and higher order $\alpha'$ corrections \cite{F4}, etc. Hence it is natural to have a potential for a generic modulus $\tau$ different from $\vo$ of the form: 
\be
V=A-\frac{B}{\tau^a}+ \cdots
\ee
with $A, B$ and $a$ positive coefficients (assuming all other moduli have been stabilised by other effects). The $\cdots$ include terms that stabilise $\tau$ but are subdominant in the inflationary regime. This is the situation in Fibre Inflation and we believe it to be more general.

The natural geometric variables determining the K\"ahler moduli are the $2$-cycle coordinates $t$ in terms of which $\vo$ and $K$ can be written explicitly as in (\ref{v}) and (\ref{K}). The idea is to use as the inflaton a canonically normalised field corresponding to a proper combination of the $\tau$ fields. Similarly to Fibre Inflation, this field is stabilised only after string loops and higher order $\alpha'$ effects are included. Given that the expression of $\vo$ is explicit in terms of the $t$ fields and only implicit in terms of the $\tau$ fields, it is simpler to work directly with the $t$ fields, identify an inflaton candidate and go to the canonically normalised field $\phi$ without the need to pass through the $\tau$ fields. We will also be interested only in the inflationary regime (not in the full potential for $\phi$). Some useful relations are:
\be
\vo_i \equiv \tau_i = \frac{\partial\vo}{\partial t_i} =\frac{1}{2} \kappa_{ijk}t_j t_k\,, \qquad 
\vo_{ij} = \frac{\partial^2\vo}{\partial t_i \partial t_j} =\kappa_{ijk}t_k\,, \qquad
\vo_{ijk}= \frac{\partial^3\vo}{\partial t_i \partial t_j \partial t_k} = \kappa_{ijk}\nn \,.
\ee
For CY manifolds, the matrix $\vo_{ij}$ has eigenvalues $(+,-,-,\cdots )$. Moreover, it is easy to verify that:
\be
\vo=\frac 13\,t_i\tau_i\,, \qquad \vo_{ij}t_j= 2\, \tau_i\,,  \qquad \vo^{-1}_{ij}\tau_j=\frac 12\, t_i\,.
\ee
The line element determining the kinetic terms is well defined for the $\tau$ variables in terms of derivatives of $K$. Based on this we can find an expression in terms of the $t$ variables:
\be
ds^2= \frac{1}{4}\frac{\partial^2 K}{\partial \tau_i\partial\tau_j} d\tau_i d\tau_j=\frac{1}{4}\left(K_{mn}-\vo_{mnp} \frac{\partial K}{\partial\tau_p}\right) dt_mdt_n\,.
\ee
Defining $Y\equiv \vo + g_s^{-3/2}\zeta/2$ and using the expressions above we can see that:
\be
K_{mn}=-2\,\frac{\vo_{mn}}{Y}+2\,\frac{\vo_m\vo_n}{Y^2}\,,\qquad 
\frac{\partial K}{\partial\tau_p}=-\frac{t_p}{Y}\,,\qquad
\vo_{mnp} t_p=\vo_{mn}\,.
\ee
Therefore we have a simpler expression for the line element $ds^2$ in terms of the $t$ fields:
\be
ds^2 =\frac 14 \left( -\frac{\vo_{mn}}{Y} +2\,\frac{\vo_m\vo_n}{Y^2}\right)dt_m dt_n\,.
\label{ds}
\ee
Given that we are interested in the inflationary regime where $\vo$ is kept constant, i.e. $d\vo=\vo_m dt_m = 0$, (\ref{ds}) symplifies to:
\be
ds^2=-\frac 14\,\frac{\vo_{mn}}{Y} \,dt_m dt_n\,.
\label{dsfin}
\ee
We concentrate on the two-field case (assuming other ones are fixed by non-perturbative effects or D-terms as in LVS constructions). In this case their differentials are related by the $d\vo=0$ condition and substituting in (\ref{dsfin}) we can determine the canonically normalised inflaton field $\phi$ for which $ds^2=\frac 12\, d\phi^2$. Notice also that in this case $\det \vo_{mn}<0$, and so the kinetic terms turn out to be correctly positive definite.
 
The most general expression for $\vo$ as a cubic on the $t$ fields in the two-field case reads:
\be
\vo= a\,t_2^3+b\,t_2^2 t_1 + c\, t_2 t_1^2\,,
\ee
where we did not write down the $t_1^3$ term since it could be generated by a simple shift like $t_2\rightarrow t_2+\gamma t_1$. Let us consider different cases:
\begin{enumerate}
\item $a=c=0$: this case corresponds to the simplest fibration where $t_1=\vo/(b\,t_2^2)$, $dt_1=-2\,t_1\, dt_2/t_2$ 
and $\vo_{11}=0$, $\vo_{12}=2 \,b\, t_2$, $\vo_{22}=2\, b\, t_1$. Hence $ds^2=-3\,b\,t_1\, dt_2^2/(2Y)\sim 3\,dt_2^2/(2\,t_2^2)=d\phi^2/2$ implying that $t_2\propto e^{\phi/\sqrt{3}}$. This is exactly what we had in Fibre Inflation for which $\tau_1\propto t_2^2=e^{2\phi/\sqrt{3}}$. 

\item $c=0$: this is the most general fibred. It is straightforward to see that we get again $t_2\propto e^{\phi/\sqrt{3}}$.
 
\item $b=0$: in this case we get:
\be
ds^2=\frac{3}{8t_2^2} \left(1-4 a \frac{t_2^3}{\vo}\right)\left(1-a \frac{t_2^3}{\vo}\right)^{-1} dt_2^2=\frac 12\,d\phi^2\,,
\ee
which can be `easily' integrated to give (with $x\equiv a\,t_2^3/\vo$):
\be
\frac{\sqrt{3}}{2} \phi= \frac{1}{4} \ln\left(\frac{\sqrt{1-x}-\sqrt{1-4x}}{\sqrt{1-x}+\sqrt{1-4x}}\right)-\cosh^{-1}\left(\frac{2}{\sqrt{3}}\sqrt{1-x}\right).
\ee
Notice that in the limit $x\rightarrow 0$ ($t_2^3\ll\vo$) this reduces to $\phi\rightarrow \frac{\sqrt{3}}{2}\ln t_2$ or $t_2\propto e^{2\phi/\sqrt{3}}$. The extra factor of $2$ in the exponential might give rise to higher values of $r$ (just a factor of $2$ more than Fibre Inflation in the best case) depending on the exact functional dependence of the inflationary potential on $t_2$. In the $x\to 0$ limit there is an exponential behaviour, and so $\epsilon\propto \eta^2$ as in the fibred case.
  
\item $a\neq 0$, $b\neq 0$, $c\neq 0$: this is the most general case. One can easily evaluate the leading terms for large $\vo$ ($\vo\gg t_2^3>1$). Solving for $t_1$ we have to leading order $t_1\simeq \sqrt{\vo/(c t_2)}$. When $\vo$ is fixed we have also: 
\be
dt_1\simeq-\frac{\vo_2}{\vo_1}\,dt_2=-\frac 12\,\sqrt{\frac{\vo}{c\,t_2^3}}\,dt_2\,.
\ee
Thus the dominant term in the metric up to errors of $\mc{O}\left(1/\sqrt{\vo}\right)$ is:
\be
ds^2  =-\frac{1}{2\vo}\left[ct_2\left(\frac{\vo_2}{\vo_1}\right)^{2}-2 (c\,t_1+b\,t_2)\frac{\vo_2}{\vo_1}\right]dt_2^2
=\frac 38\,\frac{dt_2^2}{t_2^2}=\frac 12\, d\phi^2\,.
\ee
Hence we obtain again to leading order in the large volume expansion $t_2\sim e^{2\phi/\sqrt{3}}$ which leads to an inflationary potential similar to the one of the original Fibre Inflation case. 
\end{enumerate}
In summary we have shown that the type of potential and cosmological parameters obtained in \cite{Fibre} is fairly generic in type IIB LVS string compactifications.

\subsection{After inflation}

As we have seen, in the general class of Fibre Inflation models, the moduli that play the r\^ole of the inflatons tend to be the lightest modes with $m_\phi \sim M_p/ \vo^{5/3}$. At the end of inflation, these light moduli dominate the energy density of the Universe until they settle to their minima and reheat the Universe (both observable and hidden sectors) at temperatures of order \cite{Cicoli:2010ha}:
\be
T_r \sim 0.1 \,m_\phi \sqrt{\frac{m_\phi}{M_p}} \sim 0.1 \,\frac{M_p}{\vo^{5/2}}\,.
\ee
Given that the requirement of matching the observed amplitude of the density perturbations in Fibre Inflation models leads to a relatively small volume $\vo\lsim 10^4$, all the mass scales are relatively high: the gravitino mass is very large $m_{3/2}\sim M_p/\vo\gtrsim 10^{14}$ GeV, the mass of the fibre modulus is of order $m_\phi \gtrsim 5\cdot 10^{11}$ GeV while the reheating temperature turns out to be $T_r\gtrsim 10^7$ GeV. Thus these models are safe from the cosmological moduli problem (successful Big-Bang Nucleosynthesis requires $m_\phi\gtrsim 50$ TeV \cite{CMP}) and the reheating temperature might be even large enough to allow thermal leptogenesis \cite{Lepto}. However even if the SM is sequestered from the supersymmetry breaking sector \cite{sequestered}, the superpartners are heavier than $M_{\rm soft}\sim M_p/\vo^2\gtrsim 10^{10}$ GeV. Thus these models do not allow for low-energy supersymmetry and represent a typical example of the known tension between large scale inflation and low-energy supersymmetry \cite{SUSYvsINF}.

\section{Conclusions}
\label{Concl}

The limited observational information available to cosmology limits what can be inferred about the very high energies whose physics underlies primordial fluctuations, and so satisfactory explanations should be robust enough not to depend much on these details. In this paper we point out that a class of inflationary models with this property emerges generically in string constructions as precision measurements butt up against model predictions. 
Being a generalisation of Fibre Inflation \cite{Fibre}, these models feature a simple inflationary potential with a constant term and negative exponentials, $V \simeq V_0 - V_1 \, e^{-\phi/f}$. These turn out to ($i$) describe the data well; ($ii$) do so robustly (without requirements such as detailed fine-tunings of parameters); ($iii$) arise plausibly (or, better, generically) from sensible UV completions at very high energies and ($iv$) make interesting new predictions. 

In these models the inflaton is a K\"ahler modulus different from the overall volume (as a fibration modulus for example) which enjoys an effective rescaling symmetry that protects its flatness \cite{NCNInf}. Moreover this scenario appears rather generically in string compactifications since most known Calabi-Yau manifolds feature a fibration structure \cite{GenericFibre}. 

These models robustly predict $\epsilon \simeq \eta^2$ and so also the relation $r \propto (n_s-1)^2$, where the proportionality constant is fixed by the constant $f/M_p$ of the exponentials in the inflationary potential (\ref{VBpos}). The smaller $f$ is the smaller $r$ is, and $r \simeq 0.01$ corresponds to trans-Planckian effective decay constant $f$. Our surveys of the parameter space available to Fibre Inflation models produce the comparatively narrow range $r\simeq 0.005$ to $r\simeq 0.01$, and we were unable to reach much larger (or much smaller) values. In this sense this class of models is very predictive in the sense that it can be ruled out if tensor modes are observed in the next few years but also if they are not observed in the next 10-20 years. It would be interesting to establish how generally the entire range of values for $f$ are possible in string constructions. Answering this question seems to be rather hard since $f$ could be increased or decreased by complicated mixing effects between the K\"ahler moduli \cite{Evasion}. Definitely $f$ can be very small like in K\"ahler moduli inflation \cite{kahlerinflation} where the inflaton is a small blow-up mode, but the interesting question is to understand how large it can be since large values of $r$ correspond to large values of $f$. We leave this investigation for future work.

In this paper we also compare our general inflationary scenarios with particular low-energy models which share some common features like Starobinsky $R+R^2$ inflation \cite{Starobinsky:1980te} and $\alpha$-attractors \cite{Attractors}. In Appendix \ref{AppA} we review the fact that deriving Starobinsky inflation from a UV complete theory requires the existence of at least two suppression scales:  first a scale $M \sim 10^{13}\,{\rm GeV}\ll M_p$ in order to make the $R^2$ term compete with the standard $R$ term, and a second scale which has to be larger than $M$ in order to be able to neglect higher curvature terms. However the curvature expansion obtained at low-energy from string theory is characterized by just one suppression scale which coincides with the string scale itself. We therefore concluded that it is more promising to consider models from strings by looking at scalar-tensor theories like Fibre Inflation \cite{Fibre} which do give rise to at least two different suppression scales thanks to the appearance of factors containing the VEV of the overall volume $\vo$ and the dilaton that fixes the string coupling $g_s$.

Finally, we also stressed that the self consistency of large field inflationary models -- in particular the stability of classical string-inflation scenarios against ultra-violet quantum corrections -- can give rise to strong constraints for non-supersymmetric inflationary models.

\section*{Acknowledgements}

We thank Luis Aparicio, James Gray, Renata Kallosh, Sven Krippendorf,  Andrei Linde, Anshuman Maharana, Susha Parameswaran, Gianmassimo Tasinato, Roberto Valandro, Giovanni Villadoro  and Ivonne Zavala for useful discussions. The Abdus Salam International Centre for Theoretical Physics (ICTP) kindly provided pleasant and stimulating environs while part of this work was done; a time for part of which CB also thanks the CERN TH Division for its support. This research was supported in part by funds from the Natural Sciences and Engineering Research Council (NSERC) of Canada. Research at the Perimeter Institute is supported in part by the Government of Canada through Industry Canada, and by the Province of Ontario through the Ministry of Research and Information (MRI). 

\appendix

\section{Curvature-squared models}
\label{AppA}

This appendix summarises some of the features of the Starobinsky model and its generalisations. The goal is to present it as a foil for the inflationary models described in the main text, in particular contrasting their robustness and relative difficulties with embedding them into UV completions.

\subsection{The Starobinsky model revisited}

The original Starobinsky model is based on adding a Ricci scalar squared term in addition to the standard Einstein-Hilbert action:
\be
S_s\left[g_{\mu\nu},\psi\right]=\frac{M_p^2}{2}\int d^{4}x\sqrt{-g}\left(R+\frac{R^2}{M^2} \right)+\int d^{4}x\sqrt{-g} \; L_{M}(g_{\mu\nu}.\psi),
\label{eq:Ss}
\ee
where $\psi$ represents matter fields. The gravitational part of this action is known to be equivalent to General Relativity coupled to a canonically normalised scalar field theory with a scalar potential of the type:
\be
V(\phi)=\frac 18\,M^2 M_p^2\left(1-e^{-\sqrt{\frac{2}{3}}\; \phi} \right)^2,
\ee
which has a long positive plateau leading to slow-roll inflation in the relatively large $\phi$ regime. 

Agreement with the measured amplitude of primordial density perturbations requires the coefficient $M$ be of order $M \sim 10^{-5} M_p$ (so $M \sim 10^{13}$ GeV), and leads to following predictions for inflationary observables in terms of the number of efoldings $N_e$:
\be
n_s \sim 1-\frac{2}{N_e}\qquad\text{and}\qquad r\sim \frac{12}{N_e^2}\,.
\ee
This potential is also a particular case of the more general class of exponential potentials with $k = \sqrt{2/3}$ or $\alpha = 2/3k^2 = 1$. 

The Starobinsky model and its generalisations agree well with the most recent Planck data \cite{planck15}. However the truncation of the series in $R$ and the anomalously large coefficient of $R^2$ lead one to ask what its UV provenance might be, as well as how robust a model of this type can be against quantum corrections. In particular, how might higher curvature terms (that appear naturally in any UV completion of Einstein's theory) modify the potential? We examine these issues in the next subsection, where we discuss generalisations of the model and obstacles to its possible implementation within string theory. These are to be contrasted with the story of the main text, which ask similar questions of simple inflationary models that share the observational successes of Starobinsky's model, but have clear UV origins. 

\subsection{Effects of higher curvatures}

$f(R)$ theories provide a simple generalisation of the Starobinsky model within which the model's robustness can be explored. These $f(R)$ models have the action v:
\be
S_f\left[g_{\mu\nu},\psi\right]=\frac{M_p^2}{2}\int d^{4}x\sqrt{-g}\; f(R)+\int d^{4}x\sqrt{-g}\;L_{M}(g_{\mu\nu}, \psi).\label{eq:SfR}
\ee
where $f(R)$ is supposed to be analytic around $R$=0 with $f(0)=0,\, f'(0)=1,\, f''(0)\ne 0$ and so on.

At the classical level this model is also equivalent to a scalar-tensor theory. The equivalent model is obtained by replacing the higher derivative terms by a scalar field:
\be
S_{\BD}\left[g_{\mu\nu},\psi,\chi\right] = \frac{M_p^2}{2}\int d^{4}x\sqrt{-g}\;\left[f(\chi)+f,_{\chi}(\chi)(R-\chi)\right]+\int d^{4}x\sqrt{-g}\;L_{M}(g_{\mu\nu}, \psi)\,,
\ee
so that $S_\BD$ goes to $S_f$ once $\chi$ is evaluated on shell. That is:
\be
\frac{\delta S_{\BD}}{\delta\chi} =0  \implies \chi=R \,,
\ee
at which point $S_\BD = S_f$.

This can be rewritten as a Brans-Dicke theory \cite{BD} by redefining $\varphi  :=f_{,\chi}(\chi)$ and solving for $\chi=\chi(\varphi)$
in $S_{BD}$. One finds:
\bea
S\left[g_{\mu\nu}, \psi,\varphi\right] &:=& S_{BD}\left[g_{\mu\nu}, \psi,\chi(\varphi)\right] \nn\\
&=&\int d^{4}x\sqrt{-g}\;\left[\frac{M_p^2}{2}\varphi \, R-U(\varphi)\right]+\int d^{4}x\sqrt{-g}\;L_{M}(g_{\mu\nu},\psi) \,, \label{eq:Svarphi} 
\eea
with scalar potential:
\be
 U(\varphi) = \frac{M_p^2}{2} \Bigl[ \varphi\chi(\varphi)-f(\chi(\varphi))\Bigr] \,.\label{eq:Uvarphi}
\ee
To go to Einstein frame we do a Weyl transformation: 
\bea
g_{\mu\nu}=e^{-2\omega}\tilde{g}_{\mu\nu}\,,\quad\qquad\qquad  R &= & e^{2\omega}(\tilde{R}+6\tilde{\square}\omega-6\tilde{g}^{\mu\nu}\partial_{\mu}\omega\partial_{\nu}\omega)\,, \nn \\
\tilde{S}\left[\tilde{g}_{\mu\nu}, \psi,\omega\right]\equiv S\left[e^{2\omega}\tilde{g}_{\mu\nu},\psi,\varphi=e^{2\omega}\right] & = & \int d^{4}x\sqrt{-\tilde{g}}\left[\frac{M_p^2}{2}(\tilde{R}-6\tilde{g}^{\mu\nu}\partial_{\mu}\omega\partial_{\nu}\omega) \right. \nn \\
 & &\qquad\qquad \left. \phantom{\frac12}  -e^{-4\omega}U\left(e^{2\omega}\right)+\tilde{L}_{M}(\tilde{g}_{\mu\nu},\psi)\right], \nn
\eea
and normalise by putting $\omega=\phi/\sqrt{6}$, to get:
\bea
S\left[\tilde{g}_{\mu\nu},\phi,\psi\right] & = & \tilde{S}\left[\tilde{g}_{\mu\nu}.\psi,\omega=\phi/\sqrt{6}\right]\nonumber \\
 & = & \int d^{4}x\sqrt{-\tilde{g}}\left[\frac{M_p^2}{2}\left(\tilde{R}-\tilde{g}^{\mu\nu}\partial_{\mu}\phi\partial_{\nu}\phi\right)-e^{-2\sqrt{\frac{2}{3}}\,\phi}U\left(e^{\sqrt{\frac{2}{3}}\,\phi}\right)+\tilde{L}_{M}\right].
\label{eq:Stilde}
\eea
Let us now imagine that $f(R)$ arises as a series in $R/M^2$ with $M\ll M_p$. This might be justified if the effective theory of inflation were obtained by integrating out states well below the Planck scale. Then:
\be
f(R)=R+\frac{R^2}{M^{2}}+\frac{a_{3}}{M^{4}}R^{3}+\frac{a_4}{M^6}R^{4}+\ldots.
\label{eq:fR}
\ee
Chasing through the definitions we see that the equivalent scalar theory has a potential with the expansion: 
\bea
V(\phi) &=& e^{-2\sqrt{\frac{2}{3}}\phi} U\left(e^{\sqrt{\frac{2}{3}}\phi}\right)  = \frac{M_p^2}{2}\frac{\left(\varphi\chi-f\right)}{\varphi^2} \nn\\
&=&\frac 12 M^2 M_p^2 \, e^{-2\sqrt{\frac{2}{3}}\phi}\left(U_{0}+U_{1}\, e^{\sqrt{\frac{2}{3}}\phi}+U_{2} \, e^{2\sqrt{\frac{2}{3}}\phi}+U_{3}\, e^{3\sqrt{\frac{2}{3}}\phi}+\ldots\right).
\label{eq:Vphi}
\eea
To determine the coefficients $U_i$ note the explicit transformation between $\chi$ and $\phi$ is:
\be
\varphi=e^{2\omega}=e^{\sqrt{\frac{2}{3}}\phi}=\frac{df(\chi)}{d\chi}=1+\frac{2\chi}{M^2}+\frac{3a_3}{M^4}\chi^2+\ldots \label{eq:f'phi}
\ee
Solving for $\chi$ in terms of $\phi$ and substituting in \eqref{eq:Uvarphi} gives $U$ as a series in $e^{\sqrt{\frac{2}{3}}\phi}$, whose form gives $U_i$ when compared with \pref{eq:Vphi}.  

For example consider the case with $a_{3}=0$ in \eqref{eq:fR} (such as arises in a supersymmetric theory). This gives:
\be
f(\chi(\varphi))=\frac{M^{2}}{2}(\varphi-1)+\frac{a_{4}M^2}{4} (\varphi-1)^{2}+\cdots
\ee
leading to -- after using (\ref{eq:f'phi}), (\ref{eq:Uvarphi}) and (\ref{eq:Vphi}):
\bea
V(\phi) & = & \frac 18\,M^2 M_p^2\,e^{-2\sqrt{\frac{2}{3}}\phi}\left[\left(e^{\sqrt{\frac{2}{3}}\phi}-1\right)^{2}+a_{4} \left(e^{\sqrt{\frac{2}{3}}\phi}-1\right)^{3}+\dots\right] \nn \\
 & = & \frac 18\,M^2 M_p^2\left[\left(1-e^{-\sqrt{\frac{2}{3}}\phi}\right)^{2}+ a_{4} \, e^{\sqrt{\frac{2}{3}}\phi}+\dots\right] \,.
\label{Staro}
\eea
The Starobinsky model corresponds to the special case $a_n=0$ for $n>2$. Two things are noteworthy about this expression:
\begin{itemize}
\item First, because $f(R)$ only involves a single scale $M$, one finds (and expects) the higher-order coefficients to be non-zero and $\cO(1)$.

\item  Notice that the long flat plateau at large $\phi$ is only a property of the Starobinsky case $a_n=0$ for $n> 2$. More generally positive exponentials appear, reflecting that the series expression is best-behaved for large negative $\phi$, and is ill-behaved when $\phi$ is large and positive \footnote{Generically for $f(\chi)=\chi+a_2\chi^2+\cdots$ the scalar potential $V(\phi)$ has a minimum with zero vacuum energy at $\chi=\phi=0$ ($\varphi=1$),   one maximum or inflection point for finite value of $\phi$ and then it asymptots to zero at $\varphi, \phi\rightarrow \infty$ or instead of  a maximum an inflection point. Except for the Starobinsky case for which $a_n=0, n> 2$ for which it asymptots to a positive constant, explaining the flatness of the potential in that case. This can be directly seen at least for polynomial $f$ by solving  $\partial V/\partial \phi=\sqrt{2/3}(2f-\varphi\chi)/\varphi^2=0$ (see also for instance \cite{wong} for a particular discussion). In the generic case the Starobinsky plateau is replaced by the maximum/inflection point at finite $\phi$ and depending on the flatness around this point inflation is realised or not. This explains the need of having the coefficients $a_n, n> 2$ hierarchically smaller than $a_2$. The expansion in positive exponentials is  only valid around the region $\varphi=1$ and should not be used in the region close to the maximum. The large  $\phi$ behaviour is usually different from the fibre case discussed in the text which is another way to differentiate the models.}. The inflationary asymptotics of the Starobinsky model seem not to be robust enough to survive higher corrections in higher powers of $R$. 
\end{itemize}

The breakdown of the expansion of $U(\phi)$ of the second bullet point is closely related to the well-known  \cite{Simon} breakdown of the series in $R$ at the inflationary saddle point, for which $R/M^2 \sim \mc{O}(1)$. This is the generic situation: any UV physics that provides only a single scale in the low-energy derivative expansion cannot support controlled solutions whose existence requires balancing different terms of this expansion against one another. 

Because of this, proponents of this model hope instead that the UV physics delivers a curvature expansion more along the schematic lines of:
\be
 f(R) = R + \frac{R^2}{M^2} \left[ 1 +  \frac{R}{M_p^2} + \cdots \right] \,,
\ee
wherein it is {\em only} the second term of the series that is enhanced relative to, say, the Planck scale. Although this seems a self-consistent choice there is no evidence that UV physics with the required properties exists.

\subsection{Higher curvatures from string theory}

String theory provides a concrete example of UV physics that generates curvature expansions as part of its low-energy EFT, many of whose coefficients have been computed. One can use this as a laboratory to explore for systematics amongst the coefficients of this curvature expansion.

For these purposes consider a string theory compactified on a six dimensional internal space $X$. We then assume that the volume of $X$ and the dilaton to be essentially stabilised (or at least only vary slowly) during inflation. In effect we are considering models where the inflaton is some combination of K\"ahler moduli, other than the volume.  In general one finds higher curvature terms involving the Riemann and Ricci tensor as well as the Ricci scalar, in accord with the general EFT expectation to find all interactions consistent with symmetries and particle content. 

The leading curvature terms in the effective action for 10D strings below the string scale turn out to be given schematically (in string frame) by:
\bea
I_s & = & \frac{1}{2\kappa_{10}^{2}}\int d^{10}x\sqrt{-g_{s}}\;e^{-2\Phi}\left[g_{(s)}^{MN}R_{MN}^{(s)}+(\delta_{sH}+\delta_{sI})\frac{1}{4}\alpha'g_{(s)}^{MM'}g_{(s)}^{NN'}R_{MN\,\,\, Q}^{(s)\,\, P}R_{M'N'\,\,\, P}^{(s)\,\, Q}\right.\nonumber \\
 &  & \left.+\alpha'^{3}t^{\ldots}``R^{4}"+O(\alpha'^{4})+\ldots\right] \,,\label{eq:I10}
\eea
where the $R^{2}$ terms are not present at all for IIB strings but do exist for Heterotic and type I strings (with respective labels $s=H,I$). The $\alpha'^{3}$ term is a complicated combination of four curvatures with $t^{\ldots}$ a tensor constructed out of the metric tensor (whose detailed form is not required here). Also $\kappa_{10}^{2}=(2\pi)^{7}\alpha'^{4}$ and the inverse string tension in the string frame is $\ell_s^2 = 1/M_{s}^{2}=2\pi\alpha'$. 

The metric on the internal manifold $X$ is scaled such that $g_{ij}=\vo^{1/3}\hat{g}_{ij}$
with the latter metric normalised such that $\int_{X}\sqrt{\det\hat{g}}=(2\pi)^{6}\alpha'^{3}$, so in the $\hat{g}$ metric we expect the internal space curvature $``R_{\ssX}"\sim1/\alpha'$. Using these relations and the above metrics
in \eqref{eq:I10} and keeping explicitly only terms involving the 4D curvature we have (schematically) the following structure (with
all curvatures being 4-curvatures): 
\bea
I_s & = & \frac{M_p^2}{2}\int d^{4}x\sqrt{-g}\left[ R+\ell_s^{2}``R^{2}"\left((\delta_{sH}+\delta_{sI})c_{2}+\frac{c_{4}'}{{\cal V}^{2/3}}+O\left(\frac{1}{\vo}\right)\right) \right.\nn \\
 &  & \qquad\qquad \left. +\ell_s^{6}``R^{4}" \left(c_{4}+\frac{c_{6}'}{{\cal V}^{1/3}}+O\left(\frac{1}{{\cal V}^{2/3}}\right) \right)+\ldots\right]\label{eq:4DI}
\eea
We assume the dilaton and the volume ${\cal V}$ are essentially stabilised with $e^\Phi\rightarrow g_s$ and we identify the 4D gravitational constant $M_p^{-2}=2\pi\alpha'g_s^2$. The coefficients $c_i$ are expected to be of $\mc{O}(1)$. Note that in each line the 4D curvature raised to some power comes multiplied
with a series in inverse powers of $\vo$. Several comments are in order here:
\begin{itemize}
\item Firstly the curvatures do not just come as powers of the Ricci scalar
as in $f(R)$ gravity. The leading term in the first line is actually a square of the Riemann tensor and exists only in heterotic and type
I theories ($s=H,I$). In four dimensions one could however re-express this in terms of $R^{2}$ and $R_{\mu\nu}^{2}$ and the (topological) Euler density. The $R_{\mu\nu}^{2}$ term can be removed by a field redefinition
$g_{\mu\nu}\rightarrow g_{\mu\nu}+\lambda R_{\mu\nu}$. 

\item Only one scale controls successive terms of the curvature expansion and this scale is of course the string scale. The
scale $M$ of the previous section turns out to be $M_{s}=1/\ell_s$. For the strings for which curvature-squared terms do arise (weakly coupled Type I and heterotic string theories) this scale is expected to be at most an order of magnitude or so below the Planck scale. Thus even in the best-case scenario where the leading $R^{2}$ term in the first line of \eqref{eq:4DI} is present it appears with a scale far too high to agree with the phenomenological requirement $M \sim 10^{13}$ GeV of Starobinsky-type models.

\item In type IIB theory the first $``R^{2}"$ term is absent so in the 10D theory the next-to-leading curvature term has four powers of curvature.
So the leading term beyond the Einstein term is suppressed by a scale larger than the string scale by a factor $\vo^{2/3}$ (the $c'_{4}$ term in \eqref{eq:4DI}). Also in this case the volume factor can be much larger than in the heterotic case with $\vo\gtrsim10^{3}$. 
\end{itemize}

Although we highlight here the popular Starobinsky model, we believe its robustness and difficulty embedding into a UV completion to be representative of a wider class of models. It also shares with many the difficulty with finding a UV completion, though such a completion might yet be found. If so we believe it is more fruitful to make this search regarding it as a scalar-tensor theory than from the initial formulation as a curvature expansion. 

An interesting example of a scalar-tensor theory which features more than one suppression scale is actually Fibre Inflation \cite{Fibre} whose potential (\ref{VBpos}) looks very similar to the potential (\ref{Staro}) of the Starobinsky-like model: 
\be
f(R)=R+\frac{R^2}{M^2}+\frac{a_4}{M^6} R^4+\ldots,
\ee
where $a_3=0$ is a supersymmetric theory. In fact, the similarity between the two inflationary potentials becomes very clear when we rewrite (\ref{Staro}) as:
\be
V = \frac 18\,M^2 M_p^2\left(1-2\,e^{-k\phi/\sqrt{2}} + e^{-\sqrt{2} k\phi} + a_4 \, e^{k \phi/\sqrt{2}} \right) \qquad\text{with}\qquad k=\frac{2}{\sqrt{3}}\,,
\ee
and (\ref{VBpos}) as:
\be
V = \frac 14\, M^2 M_p^2 \left(3 - 4\, e^{- k\phi/2}  + e^{- 2 k \phi} + a_4 \,e^{k \phi}\right)\,,
\ee
where:
\be
M = m_\phi \sim \frac{M_p}{\vo^{5/3}}\ll M_p\,\qquad\text{and}\qquad a_4 = R\propto g_s^4 \ll 1\,.
\ee
Hence Fibre Inflation provides a promising example of a scalar-tensor theory where one can obtain ($i$) a first suppression scale much smaller than $M_p$ due to the VEV of the internal volume $\vo$, and ($ii$) and second suppression scale which is even smaller thanks to the VEV of the dilaton which should fix the string coupling $g_s$ in the perturbative regime $g_s\ll 1$.

\section{A CY deformation of a toroidal model}
\label{AppB}

In this appendix we provide another example which satisfies the relation (\ref{rns}) between $r$ and $n_s$ but leads to a larger value of $r$ due to a smaller effective value of $k$. We start by considering the $3$-moduli $N=1$ toroidal orientifold ($T^6/(\mathbb{Z}_2\times \mathbb{Z}_2)$) where an explicit string loop computation has actually been performed (in the case of vanishing gauge fluxes) and the CY volume looks like \cite{BHK}:
\be
\vo = \sqrt{\tau_1\tau_2\tau_3}\,.
\ee
We now assume that it is possible to deform this toroidal model going to a smooth CY case by blowing up orbifold singularities. The resulting CY
volume would schematically look like:
\be
\vo = \sqrt{\tau_1\tau_2\tau_3} - \tau_s^{3/2}\,.
\ee
The addition of a blow-up mode supporting non-perturbative effects is crucial to satisfy the necessary condition to allow a standard LVS stabilisation procedure \cite{GeneralLVS}. Moduli stabilisation would schematically proceed as follows:
\begin{enumerate}
\item D-terms could fix $\tau_2$ in terms of $\tau_1$;
\item The interplay between $\alpha'$ and non-perturbative effects fix $\vo$ and $\tau_s$ following the standard LVS procedure;
\item The remaining flat direction, for example $\tau_1$, can be lifted by loop effects which develop the inflationary potential. 
\end{enumerate}
Let us describe how $\tau_2$ could be fixed by D-terms in terms of $\tau_1$: $\tau_2=\alpha\tau_1$ with $\alpha$ a positive coefficient. Non-zero gauge fluxes induce moduli-dependent Fayet-Iliopoulos (FI) terms of the form \cite{FI}:
\be
\xi_i = -q_{ij} \frac{\partial K}{\partial T_j}\,,
\ee
where $q_{ij}$ is the flux-dependent $i$-th $U(1)$-charge of the $j$-th modulus. Hence in our case we might have an FI-term that depends on both $\tau_1$ and $\tau_2$ as:
\be
\xi_i = - q_{i1} \frac{\partial K}{\partial T_1}-q_{i2} \frac{\partial K}{\partial T_2}
= \frac12 \left( \frac{q_{i1}}{\tau_1}+ \frac{q_{i2}}{\tau_2}\right),
\ee
which can fix $\tau_2=\alpha\tau_1$ by imposing $\xi_i=0$ if the two $U(1)$-charges have an opposite sign and no matter fields get a non-vanishing VEV.
Let us stress that this is just a sketchy description of a possible fixing mechanism and to check its viability in detail one would need to build a concrete model with explicit gauge fluxes. 

Let us now focus on the kinetic terms which read:
\be
\mc{L}_{\rm kin}= -\frac14 \sum_{i=1}^3 \frac{(\partial \tau_i)^2}{\tau_i^2}\,,
\label{Lkin}
\ee
together with:
\be
\frac{\partial(\vo^2)}{\vo^2} = \sum_{i=1}^3 \frac{\partial \tau_i}{\tau_i}\,.
\label{dVol}
\ee
Given that the volume is kept stable during the inflationary dynamics, (\ref{dVol}) implies:
\be
\frac{\partial \tau_3}{\tau_3}=-\left(\frac{\partial \tau_1}{\tau_1}+\frac{\partial \tau_2}{\tau_2}\right)\,.
\ee
Using now the fact that $\tau_2$ is fixed at $\tau_2=\alpha\tau_1$, we have $\partial \tau_2/\tau_2=\partial\tau_1/\tau_1$ and (\ref{Lkin}) becomes:
\be
\mc{L}_{\rm kin}= -\frac32 \frac{(\partial \tau_1)^2}{\tau_1^2}\,.
\ee
Canonical normalisation now gives:
\be
\tau_1=e^{k\phi}\qquad\text{with} \qquad k=\frac{1}{\sqrt{3}}\,,
\ee
and so now $k$ is indeed smaller by $1/2$. Focusing on the case with $R=2.25\cdot 10^{-5}$, this new value of $k$ would give:
\bea
N_e&\simeq& 50 \,: \quad n_s \simeq 0.967\qquad\text{and}\qquad r \simeq 0.019\,, \nn \\
N_e &\simeq& 60 \,:\quad n_s \simeq 0.973\qquad\text{and}\qquad r \simeq 0.015\,. \nn
\eea
The spectral index is in perfect agreement with Planck 2015 data \cite{planck15} and the tensor-to-scalar ratio turns out to be observably large.


\begin{thebibliography}{99}

\bibitem{wmap}
  E.~Komatsu {\it et al.}  [WMAP Collaboration],
  ``Seven-Year Wilkinson Microwave Anisotropy Probe (WMAP) Observations: Cosmological Interpretation,''
  Astrophys.\ J.\ Suppl.\  {\bf 192} (2011) 18
  [arXiv:1001.4538 [astro-ph.CO]].

\bibitem{planck15}
  P.~A.~R.~Ade {\it et al.} [Planck Collaboration],
  ``Planck 2015 results. XX. Constraints on inflation,''
  arXiv:1502.02114 [astro-ph.CO].

\bibitem{chaotic}
A.~D.~Linde,
  ``Chaotic Inflation,''
  Phys.\ Lett.\ B {\bf 129} (1983) 177.
  doi:10.1016/0370-2693(83)90837-7

\bibitem{NInf}
 K.~Freese, J.~A.~Frieman and A.~V.~Olinto,
  ``Natural inflation with pseudo - Nambu-Goldstone bosons,''
  Phys.\ Rev.\ Lett.\  {\bf 65} (1990) 3233.
  doi:10.1103/PhysRevLett.65.3233

  \bibitem{ExpPot}
  C.~P.~Burgess, P.~Martineau, F.~Quevedo, G.~Rajesh and R.~J.~Zhang,
  ``Brane - anti-brane inflation in orbifold and orientifold models,''
  JHEP {\bf 0203} (2002) 052
  doi:10.1088/1126-6708/2002/03/052
  [hep-th/0111025].


\bibitem{SInfPP}
  C.~P.~Burgess, M.~Cicoli and F.~Quevedo,
  ``String Inflation After Planck 2013,''
  JCAP {\bf 1311} (2013) 003
  doi:10.1088/1475-7516/2013/11/003
  [arXiv:1306.3512 [hep-th]].

\bibitem{Cicoli:2011zz}
  M.~Cicoli and F.~Quevedo,
  ``String moduli inflation: An overview,''
  Class.\ Quant.\ Grav.\  {\bf 28} (2011) 204001
  doi:10.1088/0264-9381/28/20/204001
  [arXiv:1108.2659 [hep-th]].


\bibitem{Attractors}
 R.~Kallosh and A.~Linde,
  ``Non-minimal Inflationary Attractors,''
  JCAP {\bf 1310} (2013) 033
  doi:10.1088/1475-7516/2013/10/033
  [arXiv:1307.7938 [hep-th]];

 R.~Kallosh and A.~Linde,
  ``Escher in the Sky,''
  arXiv:1503.06785 [hep-th].

\bibitem{NCNInf}
   C.~P.~Burgess, M.~Cicoli, F.~Quevedo and M.~Williams,
  ``Inflating with Large Effective Fields,''
  JCAP {\bf 1411} (2014) 045
  doi:10.1088/1475-7516/2014/11/045
  [arXiv:1404.6236 [hep-th]];

\bibitem{EInf}
 J.~Martin, C.~Ringeval and V.~Vennin,
  ``Encyclop¾dia Inflationaris,''
  Phys.\ Dark Univ.\  {\bf 5-6} (2014) 75
  doi:10.1016/j.dark.2014.01.003
  [arXiv:1303.3787 [astro-ph.CO]];
 
  J.~Martin, C.~Ringeval, R.~Trotta and V.~Vennin,
  ``The Best Inflationary Models After Planck,''
  JCAP {\bf 1403} (2014) 039
  doi:10.1088/1475-7516/2014/03/039
  [arXiv:1312.3529 [astro-ph.CO]].
  
\bibitem{CCrev}
  C.~P.~Burgess,
  ``The Cosmological Constant Problem: Why it's hard to get Dark Energy from Micro-physics,''
  doi:10.1093/acprof:oso/9780198728856.003.0004
  arXiv:1309.4133 [hep-th].

 \bibitem{GKP}
   S.~B.~Giddings, S.~Kachru and J.~Polchinski,
  ``Hierarchies from fluxes in string compactifications,''
  Phys.\ Rev.\ D {\bf 66} (2002) 106006
  doi:10.1103/PhysRevD.66.106006
  [hep-th/0105097];
  
\bibitem{KKLT}
S.~Kachru, R.~Kallosh, A.~D.~Linde and S.~P.~Trivedi,
  ``De Sitter vacua in string theory,''
  Phys.\ Rev.\ D {\bf 68} (2003) 046005
  [hep-th/0301240];
  
\bibitem{KKLMMT}
  S.~Kachru, R.~Kallosh, A.~D.~Linde, J.~M.~Maldacena, L.~P.~McAllister and S.~P.~Trivedi,
  ``Towards inflation in string theory,''
  JCAP {\bf 0310} (2003) 013
  doi:10.1088/1475-7516/2003/10/013
  [hep-th/0308055].
 
 \bibitem{LVS}
  V.~Balasubramanian, P.~Berglund, J.~P.~Conlon and F.~Quevedo,
  ``Systematics of moduli stabilisation in Calabi-Yau flux compactifications,''
  JHEP {\bf 0503} (2005) 007
  [hep-th/0502058];


 
\bibitem{eva}
  E.~Silverstein and A.~Westphal,
 ``Monodromy in the CMB: Gravity Waves and String Inflation,''
  Phys.\ Rev.\ D {\bf 78} (2008) 106003
  [arXiv:0803.3085 [hep-th]].
  
  L.~McAllister, E.~Silverstein and A.~Westphal,
  ``Gravity Waves and Linear Inflation from Axion Monodromy,''
  Phys.\ Rev.\ D {\bf 82} (2010) 046003
  [arXiv:0808.0706 [hep-th]].
  
  For a recent review with updated references see
  E.~Silverstein,
  ``Inflation in string theory confronts data,''
  Comptes Rendus Physique {\bf 16} (2015) 1003.
  doi:10.1016/j.crhy.2015.08.006
 
\bibitem{Fibre}
 M.~Cicoli, C.~P.~Burgess and F.~Quevedo,
  ``Fibre Inflation: Observable Gravity Waves from IIB String Compactifications,''
  JCAP {\bf 0903} (2009) 013
  [arXiv:0808.0691 [hep-th]].

\bibitem{kahlerinflation}
J.~P.~Conlon and F.~Quevedo,
  ``Kahler moduli inflation,''
  JHEP {\bf 0601} (2006) 146
  doi:10.1088/1126-6708/2006/01/146
  [hep-th/0509012].

\bibitem{polyinstanton}
M.~Cicoli, F.~G.~Pedro and G.~Tasinato,
  ``Poly-instanton Inflation,''
  JCAP {\bf 1112} (2011) 022
  doi:10.1088/1475-7516/2011/12/022
  [arXiv:1110.6182 [hep-th]].

 \bibitem{HI}
 F.~L.~Bezrukov and M.~Shaposhnikov,
  ``The Standard Model Higgs boson as the inflaton,''
  Phys.\ Lett.\ B {\bf 659} (2008) 703
  doi:10.1016/j.physletb.2007.11.072
  [arXiv:0710.3755 [hep-th]].

\bibitem{Starobinsky:1980te}
  A.~A.~Starobinsky,
  ``A New Type of Isotropic Cosmological Models Without Singularity,''
  Phys.\ Lett.\ B {\bf 91} (1980) 99.

\bibitem{pGB}
  S.~Weinberg, ``Nonlinear realizations of chiral symmetry,'' Phys.\ Rev.\  {\bf 166} (1968) 1568;
  doi:10.1103/PhysRev.166.1568

  S. R. Coleman, J. Wess and B. Zumino, Structure of phenomenological Lagrangians. 1., Phys. Rev. 177 (1969) 2239;
  
  C. G. Callan, Jr., S. R. Coleman, J. Wess and B. Zumino, Structure of phenomenological Lagrangians. 2., Phys. Rev. 177 (1969) 2247;
  
   S.~Weinberg, ``Approximate symmetries and pseudoGoldstone bosons,''
  Phys.\ Rev.\ Lett.\  {\bf 29} (1972) 1698;
  doi:10.1103/PhysRevLett.29.1698;
  
  S.~Weinberg, ``Phenomenological Lagrangians,''
  Physica A {\bf 96} (1979) 327.
  
\bibitem{pGBrev}
 For a review see, for instance: 
 
 C.~P.~Burgess,
  ``Goldstone and pseudoGoldstone bosons in nuclear, particle and condensed matter physics,''
  Phys.\ Rept.\  {\bf 330} (2000) 193
  doi:10.1016/S0370-1573(99)00111-8
  [hep-th/9808176].


\bibitem{GonLin}
  A.~S.~Goncharov and A.~D.~Linde,
  ``A Simple Realization Of The Inflationary Universe Scenario In Su(1,1) Supergravity,''
  Class.\ Quant.\ Grav.\  {\bf 1} (1984) L75.
  doi:10.1088/0264-9381/1/6/004


\bibitem{CKST}
   C.~Csaki, N.~Kaloper, J.~Serra and J.~Terning,
  ``Inflation from Broken Scale Invariance,''
  Phys.\ Rev.\ Lett.\  {\bf 113} (2014) 161302
  doi:10.1103/PhysRevLett.113.161302
  [arXiv:1406.5192 [hep-th]].

\bibitem{GenericFibre}
  P.~Candelas, A.~Constantin and H.~Skarke,
  ``An Abundance of K3 Fibrations from Polyhedra with Interchangeable Parts,''
  Commun.\  Math.\  Phys.\  {\bf 324} (2013) 937
  doi:10.1007/s00220-013-1802-2
  [arXiv:1207.4792 [hep-th]];
  L.~B.~Anderson, F.~Apruzzi, X.~Gao, J.~Gray and S.~J.~Lee,
  ``A New Construction of Calabi-Yau Manifolds: Generalized CICYs,''
  arXiv:1507.03235 [hep-th].
 
 \bibitem{BHK}
   M.~Berg, M.~Haack and B.~Kors,
  ``String loop corrections to Kahler potentials in orientifolds,''
  JHEP {\bf 0511} (2005) 030
  doi:10.1088/1126-6708/2005/11/030
  [hep-th/0508043].
  
 \bibitem{HiDInf}
 L.~van Nierop and C.~P.~Burgess,
  ``Sculpting the Extra Dimensions: Inflation from Codimension-2 Brane Back-reaction,''
  JCAP {\bf 1204} (2012) 037
  doi:10.1088/1475-7516/2012/04/037
  [arXiv:1108.2553 [hep-th]].

   \bibitem{NonAdiabEFT}
   C.~P.~Burgess,
  ``Quantum gravity in everyday life: General relativity as an effective field theory,''
  Living Rev.\ Rel.\  {\bf 7} (2004) 5
  doi:10.12942/lrr-2004-5
  [gr-qc/0311082];
  
   C.~P.~Burgess and M.~Williams,
  ``Who You Gonna Call? Runaway Ghosts, Higher Derivatives and Time-Dependence in EFTs,''
  JHEP {\bf 1408} (2014) 074
  doi:10.1007/JHEP08(2014)074
  [arXiv:1404.2236 [gr-qc]].

  \bibitem{GSnogo}
  T. Banks and L. J. Dixon, Nucl. Phys. B 307 (1988) 93.   
 
   \bibitem{GSgo}
C.~P.~Burgess, J.~P.~Conlon, L.~Y.~Hung, C.~H.~Kom, A.~Maharana and F.~Quevedo,
  ``Continuous Global Symmetries and Hyperweak Interactions in String Compactifications,''
  JHEP {\bf 0807} (2008) 073
  doi:10.1088/1126-6708/2008/07/073
  [arXiv:0805.4037 [hep-th]].

\bibitem{LiamEva}
  L.~McAllister and E.~Silverstein,
  ``String Cosmology: A Review,''
  Gen.\ Rel.\ Grav.\  {\bf 40} (2008) 565
  doi:10.1007/s10714-007-0556-6
  [arXiv:0710.2951 [hep-th]].

  \bibitem{NonAdiabExamples}
  C.~P.~Burgess, J.~M.~Cline, F.~Lemieux and R.~Holman,
  ``Are inflationary predictions sensitive to very high-energy physics?,''
  JHEP {\bf 0302} (2003) 048
  doi:10.1088/1126-6708/2003/02/048
  [hep-th/0210233];
  

 \bibitem{InfEFT}
    C.~P.~Burgess, J.~M.~Cline and R.~Holman,
  ``Effective field theories and inflation,''
  JCAP {\bf 0310} (2003) 004
  doi:10.1088/1475-7516/2003/10/004
  [hep-th/0306079].
  
 \bibitem{HoverM}
N.~Kaloper, M.~Kleban, A.~E.~Lawrence and S.~Shenker,
  ``Signatures of short distance physics in the cosmic microwave background,''
  Phys.\ Rev.\ D {\bf 66} (2002) 123510
  doi:10.1103/PhysRevD.66.123510
  [hep-th/0201158].
    
  
   \bibitem{DispRel}
   
   For a review see
   
   R.~H.~Brandenberger and J.~Martin,
  ``Trans-Planckian Issues for Inflationary Cosmology,''
  Class.\ Quant.\ Grav.\  {\bf 30} (2013) 113001
  doi:10.1088/0264-9381/30/11/113001
  [arXiv:1211.6753 [astro-ph.CO]].

\bibitem{Lyth}
 D.~H.~Lyth,
  ``What would we learn by detecting a gravitational wave signal in the cosmic microwave background anisotropy?,''
  Phys.\ Rev.\ Lett.\  {\bf 78} (1997) 1861
  doi:10.1103/PhysRevLett.78.1861
  [hep-ph/9606387].
 /03/052;
 
\bibitem{Pramod}
  A.~Mazumdar and P.~Shukla,
  ``Model independent bounds on tensor modes and stringy parameters from CMB,''
  arXiv:1411.4636 [hep-th].

J.~P.~Conlon, F.~Quevedo and K.~Suruliz,
  ``Large-volume flux compactifications: Moduli spectrum and D3/D7 soft supersymmetry breaking,''
  JHEP {\bf 0508} (2005) 007
  [hep-th/0505076].
  
\bibitem{BHP}
  M.~Berg, M.~Haack and E.~Pajer,
  ``Jumping Through Loops: On Soft Terms from Large Volume Compactifications,''
  JHEP {\bf 0709} (2007) 031
  doi:10.1088/1126-6708/2007/09/031
  [arXiv:0704.0737 [hep-th]].
	
\bibitem{Loop}
	M.~Cicoli, J.~P.~Conlon and F.~Quevedo,
  ``Systematics of String Loop Corrections in Type IIB Calabi-Yau Flux Compactifications,''
  JHEP {\bf 0801} (2008) 052
  doi:10.1088/1126-6708/2008/01/052
  [arXiv:0708.1873 [hep-th]].
	
\bibitem{Math}	
K.~Oguiso, 
``On Algebraic Fiber Space Structures on a Calabi-Yau 3-fold,'' 
Int. J. of Math. {\bf 4} (1993) 439-465;

M.~B.~Schulz,
  ``Calabi-Yau duals of torus orientifolds,''
  JHEP {\bf 0605} (2006) 023
  doi:10.1088/1126-6708/2006/05/023
  [hep-th/0412270].
	
\bibitem{CYmodels}	
M.~Cicoli, M.~Kreuzer and C.~Mayrhofer,
  ``Toric K3-Fibred Calabi-Yau Manifolds with del Pezzo Divisors for String Compactifications,''
  JHEP {\bf 1202} (2012) 002
  doi:10.1007/JHEP02(2012)002
  [arXiv:1107.0383 [hep-th]].
	
\bibitem{new}
R.~Kallosh and T.~Wrase,
  ``Emergence of Spontaneously Broken Supersymmetry on an Anti-D3-Brane in KKLT dS Vacua,''
  JHEP {\bf 1412} (2014) 117
  doi:10.1007/JHEP12(2014)117
  [arXiv:1411.1121 [hep-th]];
  
E.~A.~Bergshoeff, K.~Dasgupta, R.~Kallosh, A.~Van Proeyen and T.~Wrase,
  ``$ \overline{\mathrm{D}3} $ and dS,''
  JHEP {\bf 1505} (2015) 058
  doi:10.1007/JHEP05(2015)058
  [arXiv:1502.07627 [hep-th]];
  
  R.~Kallosh, F.~Quevedo and A.~M.~Uranga,
  ``String Theory Realizations of the Nilpotent Goldstino,''
  JHEP {\bf 1512} (2015) 039
  doi:10.1007/JHEP12(2015)039
  [arXiv:1507.07556 [hep-th]];
  
	L.~Aparicio, F.~Quevedo and R.~Valandro,
  ``Moduli Stabilisation with Nilpotent Goldstino: Vacuum Structure and SUSY Breaking,''
  JHEP {\bf 1603} (2016) 036
  doi:10.1007/JHEP03(2016)036
  [arXiv:1511.08105 [hep-th]];
  
	I.~Garcia-Etxebarria, F.~Quevedo and R.~Valandro,
  ``Global String Embeddings for the Nilpotent Goldstino,''
  JHEP {\bf 1602} (2016) 148
  doi:10.1007/JHEP02(2016)148
  [arXiv:1512.06926 [hep-th]].
	
\bibitem{Tbranes}	
M.~Cicoli, F.~Quevedo and R.~Valandro,
  ``De Sitter from T-branes,''
  arXiv:1512.04558 [hep-th].
	
\bibitem{NPD3s}	
M.~Cicoli, A.~Maharana, F.~Quevedo and C.~P.~Burgess,
  ``De Sitter String Vacua from Dilaton-dependent Non-perturbative Effects,''
  JHEP {\bf 1206} (2012) 011
  doi:10.1007/JHEP06(2012)011
  [arXiv:1203.1750 [hep-th]].

\bibitem{GeneralLVS}
  M.~Cicoli, J.~P.~Conlon and F.~Quevedo,
  ``General Analysis of LARGE Volume Scenarios with String Loop Moduli Stabilisation,''
  JHEP {\bf 0810} (2008) 105
  doi:10.1088/1126-6708/2008/10/105
  [arXiv:0805.1029 [hep-th]].

\bibitem{F4}
  D.~Ciupke, J.~Louis and A.~Westphal,
  ``Higher-Derivative Supergravity and Moduli Stabilization,''
  JHEP {\bf 1510} (2015) 094
  doi:10.1007/JHEP10(2015)094
  [arXiv:1505.03092 [hep-th]].

 \bibitem{Broy:2015zba}
B.~J.~Broy, D.~Ciupke, F.~G.~Pedro and A.~Westphal,``Starobinsky-Type Inflation from $\alpha'$-Corrections,''
 JCAP01(2016)001doi:10.1088/1475-7516/2016/01/001[arXiv:1509.00024 [hep-th]].

\bibitem{ckl}
  J.~J.~M.~Carrasco, R.~Kallosh and A.~Linde,
  ``$\alpha $-Attractors: Planck, LHC and Dark Energy,''
  JHEP {\bf 1510} (2015) 147
  doi:10.1007/JHEP10(2015)147
  [arXiv:1506.01708 [hep-th]].

\bibitem{FI}
  M.~Haack, D.~Krefl, D.~Lust, A.~Van Proeyen and M.~Zagermann,
  ``Gaugino Condensates and D-terms from D7-branes,''
  JHEP {\bf 0701} (2007) 078
  doi:10.1088/1126-6708/2007/01/078
  [hep-th/0609211].

\bibitem{Cicoli:2010ha}
  M.~Cicoli and A.~Mazumdar,
  ``Reheating for Closed String Inflation,''
  JCAP {\bf 1009} (2010) 09,  025
  doi:10.1088/1475-7516/2010/09/025
  [arXiv:1005.5076 [hep-th]].

\bibitem{CMP}
G.~D.~Coughlan, W.~Fischler, E.~W.~Kolb, S.~Raby and G.~G.~Ross,
  ``Cosmological Problems for the Polonyi Potential,''
  Phys.\ Lett.\ B {\bf 131} (1983) 59;
  
T.~Banks, D.~B.~Kaplan and A.~E.~Nelson,
  ``Cosmological implications of dynamical supersymmetry breaking,''
  Phys.\ Rev.\ D {\bf 49} (1994) 779;

B.~de Carlos, J.~A.~Casas, F.~Quevedo and E.~Roulet,
  ``Model independent properties and cosmological implications of the dilaton and moduli sectors of 4-d strings,''
  Phys.\ Lett.\ B {\bf 318} (1993) 447.

\bibitem{Lepto}
M.~Fukugita and T.~Yanagida,
  ``Baryogenesis Without Grand Unification,''
  Phys.\ Lett.\ B {\bf 174} (1986) 45.
  doi:10.1016/0370-2693(86)91126-3
	
\bibitem{sequestered}
R.~Blumenhagen, J.~P.~Conlon, S.~Krippendorf, S.~Moster and F.~Quevedo,
  ``SUSY Breaking in Local String/F-Theory Models,''
  JHEP {\bf 0909} (2009) 007
  doi:10.1088/1126-6708/2009/09/007
  [arXiv:0906.3297 [hep-th]];

  S.~P.~de Alwis,
  ``Classical and Quantum SUSY Breaking Effects in IIB Local Models,''
  JHEP {\bf 1003} (2010) 078
  doi:10.1007/JHEP03(2010)078
  [arXiv:0912.2950 [hep-th]].
  
L.~Aparicio, M.~Cicoli, S.~Krippendorf, A.~Maharana, F.~Muia and F.~Quevedo,
  ``Sequestered de Sitter String Scenarios: Soft-terms,''
  JHEP {\bf 1411} (2014) 071
  doi:10.1007/JHEP11(2014)071
  [arXiv:1409.1931 [hep-th]].

\bibitem{SUSYvsINF}  
 C.~P.~Burgess, J.~M.~Cline, H.~Stoica and F.~Quevedo,
  ``Inflation in realistic D-brane models,''
  JHEP {\bf 0409} (2004) 033
  doi:10.1088/1126-6708/2004/09/033
  [hep-th/0403119];

 R.~Kallosh and A.~D.~Linde,
  ``O'kklt,''
  JHEP {\bf 0702} (2007) 002
  doi:10.1088/1126-6708/2007/02/002
  [hep-th/0611183].

\bibitem{Evasion}
  J.~E.~Kim, H.~P.~Nilles and M.~Peloso,
  ``Completing natural inflation,''
  JCAP {\bf 0501}, 005 (2005)
  [hep-ph/0409138].

  T.~C.~Bachlechner, C.~Long and L.~McAllister,
  ``Planckian Axions in String Theory,''
  arXiv:1412.1093 [hep-th].

  K.~Choi, H.~Kim and S.~Yun,
  ``Natural inflation with multiple sub-Planckian axions,''
  Phys.\ Rev.\ D {\bf 90} (2014) 023545
  [arXiv:1404.6209 [hep-th]].

  C.~Long, L.~McAllister and P.~McGuirk,
  ``Aligned Natural Inflation in String Theory,''
  Phys.\ Rev.\ D {\bf 90} (2014) 023501
  [arXiv:1404.7852 [hep-th]];
  I.~Ben-Dayan, F.~G.~Pedro and A.~Westphal,
  ``Towards Natural Inflation in String Theory,''
  arXiv:1407.2562 [hep-th].

 C.~Burgess and D.~Roest,
  ``Inflation by Alignment,''
  JCAP {\bf 1506} (2015) 06,  012
  doi:10.1088/1475-7516/2015/06/012
  [arXiv:1412.1614 [hep-th]].

\bibitem{BD}
 C. Brans and R. H. Dicke, Mach's Principle and a Relativistic Theory of Gravitation, Phys. Rev. 124 (1961) 925;
 
 R. H. Dicke, Mach's Principle and Invariance under Transformation of Units, Phys. Rev. 125 (1962) 2163;
 
 C. Brans, Mach's Principle and a Relativistic Theory of Gravitation. II, Phys. Rev. 125 (1962) 2194.

\bibitem{Simon}
  J.~Z.~Simon,
  ``No Starobinsky inflation from selfconsistent semiclassical gravity,''
  Phys.\ Rev.\ D {\bf 45} (1992) 1953.
  doi:10.1103/PhysRevD.45.1953

\bibitem{wong}
Q.~G.~Huang,
  ``A polynomial f(R) inflation model,''
  JCAP {\bf 1402} (2014) 035
  doi:10.1088/1475-7516/2014/02/035
  [arXiv:1309.3514 [hep-th]].
  
\end{thebibliography}
\end{document}